\documentclass{emulateapj}
\usepackage{txfonts}
\usepackage[figuresright]{rotating}

\shorttitle{Properties of M31 globular clusters}
\shortauthors{Wang et al.}

\begin{document}

\slugcomment{AJ, in press}
\title{STRUCTURAL PARAMETERS FOR GLOBULAR CLUSTERS IN M31}
\author{Song Wang,\altaffilmark{1,2,3}  Jun Ma\altaffilmark{1,3} }

\altaffiltext{1}{National Astronomical Observatories, Chinese Academy of Sciences, Beijing, 100012, China;\\ majun@nao.cas.cn}

\altaffiltext{2}{University of Chinese Academy of Sciences, Beijing, 100039, China}

\altaffiltext{3}{Key Laboratory of Optical Astronomy, National Astronomical Observatories, Chinese Academy of Sciences, Beijing, 100012, China}

\begin{abstract}

In this paper, we present surface brightness profiles for 79 globular clusters in M31, using images observed with {\it Hubble Space Telescope}, some of which are from new observations. The structural and dynamical parameters are derived from fitting the profiles to several different models for the first time. The results show that in the majority of cases, King models fit the M31 clusters as well as Wilson models, and better than S\'{e}rsic models. However, there are 11 clusters best fitted by S\'{e}rsic models with the S\'{e}rsic index $n>2$, meaning that they have cuspy central density profiles. These clusters may be the well-known core-collapsed candidates. There is a bimodality in the size distribution of M31 clusters
at large radii, which is different from their Galactic counterparts. In general, the properties of clusters in M31 and the Milky Way fall in the same regions of parameter spaces. The tight correlations of cluster properties indicate a ``fundamental plane'' for clusters, which
reflects some universal physical conditions and processes operating at the epoch of cluster formation.

\end{abstract}


\section{INTRODUCTION}
\label{Introduction.sec}

It is well known that, studying the spatial structures and dynamics of globular clusters (GCs) is of great importance for understanding both their formation condition and dynamical evolution within the environment of their galaxies \citep{mclau08}. For example, these clusters are ideal laboratories for detailed studies on two-body relaxation, mass segregation, stellar collisions and mergers, and core collapse \citep{mh97}. The correlations of structures with galactocentric distance can provide information on the role of the galaxy tides towards
the clusters, while the distribution of ellipticity
can shed light on the primary factor for cluster elongation. In addition, comparisons of structures of GCs
located in different environment of galaxies offer clues to differences in the early formation and evolution of the galaxies or in their subsequent accretion histories \citep{bfi,mackey07}. The ``fundamental plane'' for clusters in parameter space reflects universal cluster formation conditions, regardless of their host environments \citep{barmby09}.

The structural and dynamical parameters of clusters are often determined by fitting the surface brightness profiles to structure models, combined with mass-to-light ratios
estimated from velocity dispersions or population-synthesis models. An accurate and well resolved density profile can be obtained by studying the distribution of integrated light coupling with star counts \citep{federici07}. Several models are often used in the fits: the empirical, single-mass, modified isothermal spheres
\citep{king62,king66,wilson75}; the isotropic multi-mass models \citep{df76}; the anisotropic multi-mass models \citep{gg79,meylan88,meylan89}; the power-law surface-density profiles \citep{sersic68,elson87}.

The nearest large GC system outside the Milky Way (MW) is that of M31, with a distance of $\sim784$ kpc from us \citep{sg98}. It is so close to us that most GCs in it can be well resolved with {\it Hubble Space Telescope} ({\it HST}). \citet{battistini82} first estimated core radii for several clusters in M31, and subsequently, a number of studies
\citep{pv84,crampton85,spassova88,bendinelli90,bendinelli93,cf91,fusi94,bhh02,
ma06,ma07,ma12} focused on the internal structures of M31 GCs, including the core radius, half-light radius, tidal radius, and ellipticity, using the images from large ground-based telescopes and {\it HST}. \citet{barmby07} derived structural and dynamical parameters for 34 M31 GCs, and construct a comprehensive catalog of these parameters
for 93 M31 GCs with corrected versions of those in a previous study \citep{bhh02}. Combined with the structures and dynamics for clusters from the MW, Magellanic Clouds, the Fornax dwarf spheroidal, and NGC 5128, these authors found the GCs have near-universal structural properties,
regardless of their host environments. \citet{barmby09} found that bright young clusters in M31 are larger and more concentrated than old ones, and are expected to dissolve within a few Gyr and will not survive to become old GCs.
With measurements of structural parameters for 13 extended clusters (ECs) in the halo regions of M31, \citet{huxor11} presented that the faintest ECs have magnitudes and sizes similar to Palomar-type GCs in the MW halo.
\citet{wang12a} measured structures and kinematics for 10 newly discovered GCs in the outer halo of M31, and found that they have larger ellipticities than most of GCs in M31 and the MW, which may be due to galaxy tides from satellite galaxies of M31 or may be related to the merger and accretion history that M31 has experienced. Using the same sample clusters in \citet{wang12a}, \citet{tanvir12} found that some GCs in M31 exhibit cuspy cores which are well described by \citet{sersic68} models. These authors also confirmed the exist of luminous and compact globulars
at large galactocentric radii of M31, with no counterparts found in the MW. The last three studies extended the structural analysis of clusters in M31 out to $R_{\rm gp}\sim100$ kpc, providing important information on
the accretion history of M31 outer regions.

In this paper, we determined structures and kinematics for 79 clusters in M31 by fitting several structural models to
their surface brightness profiles. This paper is organized as follows. In Section 2, we present the {\it HST} observations for the sample clusters, and the data-processing steps to derive the surface brightness profiles. In Section 3, we determine structures and kinematics of the clusters and make some comparisons with
previous studies. A discussion on the correlations of the
derived parameters is given in Section 4. Finally, we summarize our results in Section 5.

\section{DATA AND ANALYSIS PROCEDURES}
\label{data.sec}

\subsection{Globular Cluster Sample}

{\sl HST} has imaged a large fraction of globular clusters (GCs) in M31. \citet{bhh02} used {\sl HST}/Wide Field Planetary Camera 2 (WFPC2) images to measure ellipticities, position angles, and best-fit \citet{king66} model
(hereafter ``King model'') parameters for a large sample of M31 GCs. \citet{barmby07} determined structures and kinematics for M31 GCs by fitting surface brightness profiles to different models, however, for all the GCs but G001 studied in \citet{bhh02}, only King model was used.
In addition, there were some clusters located at the edges of the images or observed with only one filter. With updated observations by {\sl HST}, new data can be derived for them now. So, we decided to re-estimate the structure parameters for these GCs in \citet{bhh02}.
However, \citet{barmby07} determined structures for G001 using three models and new observation, while \citet{barmby09} determined structure parameters for five clusters (B315, B318, B319, B368, and B374) using updated {\sl HST} data. In addition, there is no new observation for B077, which is at the edges of the {\sl HST} images.
We also noticed that BH20 and BH21 had been classified as stars, while BH24 a galaxy \citep{cald09}, which would not be discussed again in this paper. The remaining 69 clusters from \citet{bhh02} would be included in the sample. \citet{mackey07} estimated metallicities, distance moduli and reddening values for 10 newly-discovered halo GCs in M31 using the ACS/Wide Field Camera (WFC) images. Although \citet{wang12a} and \citet{tanvir12} determined structures for the 10 GCs, \citet{wang12a} only used King model, while \citet{tanvir12} presented few structure parameters. So, we re-estimated the structure parameters for these GCs. Finally, there are 79 clusters in our sample. We obtained the combined drizzled images from the Hubble Legacy Archive. The images in the bandpass close to $V$ band (F555W or F606W) and $I$ band (F814W) were preferred, otherwise the images of F300W, F435W or F475W were selected. The images with high resolution were preferentially adopted (WFPC2/PC or ACS/WFC), followed by those obtained with WFPC2/WFC. Figure 1 shows the spatial distribution for the sample GCs.

\begin{figure}
\figurenum{1} \resizebox{\hsize}{!}{\rotatebox{0}{\includegraphics{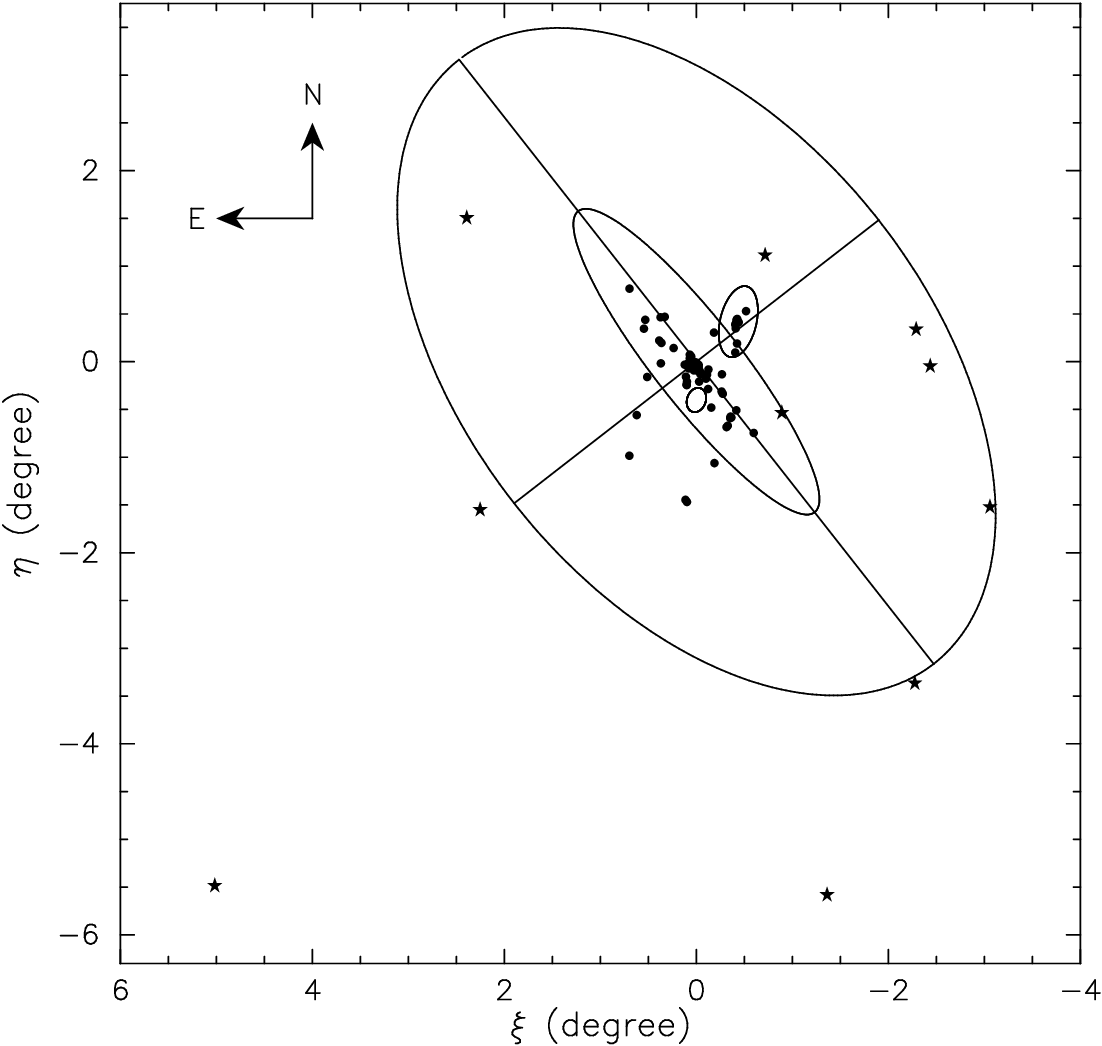}}}
\caption{Location of our sample GCs in relation to M31. The inner ellipse delineates M31's main stellar disk ($i = 77^{\circ}$ and $R =2^{\circ}$ ) while the outer ellipse has a radius of 55 kpc and is flattened to $b/a = 0.6$, as given in \citet{rich09}. The filled circles and asterisks represent the sample GCs from \citet{bhh02} and
\citet{mackey07}, respectively. The two small ellipses near the M31 center represent M32 (bottom) and NGC 205 (top-right).} \label{fig:fig1}
\end{figure}

\subsection{Surface Brightness Profiles}
\label{brightness.sec}

The data analysis procedures to measure surface brightness
profiles of clusters have been described in \citet{barmby07}. When the center positions of these clusters were determined using the {\sc imcentroid} task in IRAF, the {\sc ellipse} task was run in two passes to derive the surface brightness profiles.
The {\sc ellipse} showed inability to converge for several individual clusters. In these cases we first smoothed the images with a boxcar filter \citep{larsen02}, and then ran the {\sc ellipse} to derive the density profiles.
The overall ellipticity and position angle (PA) were determined by averaging the {\sc ellipse} output in the first pass, with the errors estimated as the standard deviation of the mean. Several clusters (B330, B468, BH04, BH11, BH29, and NB39) showed errors of P.A.s larger than 15 degrees. We checked the images for these clusters, and found that the random fluctuations due to individual stars \citep{larsen02} may account for the high errors, leading to a more difficult business for accurate measurements of PAs Table 1 lists the average ellipticity, P.A. and some additional integrated data for the sample clusters.
Considering that the metallicities for young ($<1$ Gyr) and old clusters are quite different, first, we averaged the ages for the sample 79 clusters from several previous studies \citep{cald09,cald11,kang12,wang10,wang12b}
to distinguish young from old clusters. Of which, there are 13 clusters \citep[BH23, BH29, NB39, and the 10 GCs from][]{mackey07} with no available age values in the literatures, and we assumed them as old ones. Finally, there are six young clusters (B097D, B324, BH05, BH12, DAO38, and M091) in our sample, all of which are from \citet{bhh02}. For the 69 GCs from \citet{bhh02}, the metallicities with uncertainties from \citet{kang12}
were preferentially adopted as our reference, followed by
those of \citet{cald09,cald11} for old clusters,
while the solar metallicity was assumed for young clusters
\citep{barmby09}. The reddening values were from \citet{kang12}, while the other integrated data were from RBC V.5 \citep{gall04,gall06,gall09}.
For the 10 GCs from \citet{mackey07}, we used the
integrated data presented in \citet{wang12a}, including
the $VI$ magnitudes, galactocentric distances, reddening values, and metallicities \citep[see][and references therein]{wang12a}. In addition, $B$ magnitudes for six of the 10 GCs are from RBC V.5. Old clusters with no metallicity measurements are assigned with a
mean M31 GC metallicity of [Fe/H]$=-1.2$ \citep{huchra91},
while the uncertainties of [Fe/H] are assumed to be 0.6 as for the standard deviation of the metallicity distribution of the M31 GC system \citep{bh00}. Clusters with no reddening values are assigned with the Galactic reddening in the direction of M31 of $E(B-V)=0.08$ \citep{vanden69}.

Raw output from package {\sc ellipse} is in terms
of counts s$^{-1}$ pixel$^{-1}$, which needs to multiply by a number (400 for ACS/WFC, 100 for WFPC2/WFC, and 494 for WFPC2/PC) to convert to counts s$^{-1}$ arcsec$^{-1}$.
A formula was used to transform counts to surface brightness in magnitude calibrated on the {\sc vegamag} system,

\begin{equation}
\mu/{\rm mag~arcsec^{-2}=
-2.5 \log(counts~s^{-1}~arcsec^{-1}) + Zeropoint}.
\end{equation}

As noted by \citet{barmby07}, occasional oversubtraction of background during the multidrizzling in the automatic reduction pipeline led to ``negative'' counts in some pixels, so we worked in terms of linear intensity instead of surface brightness in magnitude. With updated absolute magnitudes of the sun $M_{\odot}$ (C. Willmer, private communication) listed in Table 2, the equation for transforming counts to surface brightness in intensity was derived,

\begin{equation}
I/L_{\odot}~{\rm pc^{-2}
\simeq Conversion~Factor\times(counts~s^{-1}~arcsec^{-1})}.
\end{equation}

Converting from luminosity density in $L_{\odot}~{\rm pc^{-2}}$ to surface brightness in magnitude was done according to
\begin{equation}
\mu/{\rm mag~arcsec^{-2}}=
-2.5\log(I/L_{\odot}~{\rm pc^{-2}}) + {\rm Coefficient}.
\end{equation}

Table 2 presents the Zeropoints, Conversion Factors, and Coefficients used in these transformations for each filter. Table 3 gives the final, calibrated intensity profiles for the 79 clusters but with no extinction corrected. The reported intensities are calibrated on the {\sc vegamag} scale. Column 7 gives a flag for each point, which has the same meaning as \citet{barmby07} and \citet{mclau08} defined. The points flagged with``OK'' are used to constrain the model fit, while the points flagged with ``DEP'' are those that may lead to excessive weighting of the central regions of clusters \citep[see][for details]{barmby07,mclau08}.
In addition, points marked with ``BAD'' are those individual isophotes that deviated strongly from their neighbors or showed irregular features, which were deleted by hand.

\subsection{Point-spread Function}

The point-spread function (PSF) models are critical to accurately measure the shapes of objects in images taken with {\sl HST} \citep{rhodes06}. \citet{bhh02} found that by fitting models without PSF convolution,
the scale radii were systematically larger, and the concentrations smaller than those estimated from the convolved models. Compared to ground-based telescopes, the PSF of {\sl HST} is very stable, although it is also known to vary with time \citep{krist11}. Tiny Tim has been the standard modeling software for {\sl HST} PSF
simulation for 20 years, with a variety of uses ranging from deconvolution, model convolution, PSF fitting photometry and astrometry, and PSF subtraction \citep{krist11}. In this paper, we derived the ACS/WFC and WFPC2 PSF models using Tiny Tim \footnote{http://tinytim.stsci.edu/cgi-bin/tinytimweb.cgi.},
and then the models were fitted using a function of the form

\begin{equation}
I_{\rm PSF}/I_{\rm 0} = [1 + (R/r_0)^{\alpha}]^{-{\beta}/{\alpha}},
\label{eq:psf}
\end{equation}

where $r_0$, $\alpha$, and $\beta$ for each filter are given in Table 4. It can be seen that the parameters in Table 4 are slightly different with those from \citet{barmby07} for ACS/WFC in the F606W and F814W filters. \citet{barmby07} selected a number of isolated stars on a number of images, and combined them to produce a single, average PSF for each filter. Here we derived a few model PSFs at different positions of the camera, and averaged them to produce the final PSF for each filter. The discrepancies of these parameters from the two studies are due to the different methods, but are negligible. The PSF variation over the cluster extent was ignored since the clusters are small compared to the camera field of view \citep{barmby09}.

\section{MODEL FITTING}
\label{fit.sec}

\subsection{Structure Models}

We used three structural models to fit star cluster surface profiles, including King model, \citet{wilson75}, and \citet{sersic68} model (hereafter ``Wilson model'' and ``S\'{e}rsic model''). \citet{mclau08} have described the three structural models in detail, here we briefly summarized some basic characteristics for them.

King model is most commonly used in studies of star clusters, which is the simple model of single-mass, isotropic, modified isothermal sphere.
\citet{barmby07, barmby09} found that M31 clusters are better fitted by King models. The phase-space distribution function for King model is defined as
\begin{equation}
f(E) \propto\left\{
\begin{array}{lcl}
\exp[-E/{\sigma}_0^2]-1, &      & E < 0, \\
0,                       &      & E \geq 0,
\end{array}\right.
\end{equation}
where $E$ is the stellar energy, ${\sigma}_0$ is a velocity scale.

Wilson model is an alternate modified isothermal sphere based on the ad hoc stellar distribution function of \citet{wilson75}. These models have more extended envelope structures than the standard King isothermal spheres \citep{mclau08}. Several studies presented that Wilson models fit the majority of GCs in the Milky Way (MW) and some of its satellites and NGC 5128 as well as or better than King models \citep{mm05,mclau08}. The phase-space distribution function for Wilson model is defined as
\begin{equation}
f(E) \propto\left\{
\begin{array}{lcl}
\exp[-E/{\sigma}_0^2]-1+E/{\sigma}_0^2, &      & E < 0, \\
0,                       &      & E \geq 0.
\end{array}\right.
\end{equation}

S\'{e}rsic model has a $R^{1/n}$ surface-density profile,
and has been the standard model for parameterizing the surface brightness profiles of early-type galaxies and bulges of spiral galaxies \citep{bg11}.
\citet{tanvir12} found that some classical GCs in M31 which exhibit cuspy core profiles are well described by S\'{e}rsic models of index $n\sim2-6$. The clusters with cuspy cores have usually been called post-core collapse \citep[see][and references therein]{ng06}.
The S\'{e}rsic model is defined as
\begin{equation}
I(R)=I_0\exp[-\ln(2)\times(R/r_0)^{(1/n)}].
\end{equation}

\subsection{Fits}

Before we fitted models to the brightness profiles of the
sample clusters, the intensity profiles were corrected for extinction. Table 2 lists the effective wavelengths of the ACS and WFPC2 filters from the Instrument Handbook.
With the extinction curve taken from \citet{car89} with
$R_V=3.1$, we derived the $A_{\lambda}$ values for each filter.

We first convolved the three models with PSFs for the filters used. Given a value for the scale radius $r_0$, a dimensionless model profile $\widetilde{I}_{\rm mod}\equiv I_{\rm mod}/I_0$ was computed, and then the convolution was carried out,

\begin{equation}
\widetilde{I}_{\rm mod}^{*} (R | r_0) = \int\!\!\!\int_{-\infty}^{\infty}
               \widetilde{I}_{\rm mod}(R^\prime/r_0)
               \widetilde{I}_{\rm PSF}
               \left[(x-x^\prime),(y-y^\prime)\right]dx^\prime dy^\prime,
\label{eq:convol}
\end{equation}
where $R^2=x^2+y^2$, and $R^{\prime2}=x^{\prime2}+y^{\prime2}$.
$\widetilde{I}_{\rm PSF}$ was approximated using
the equation (\ref{eq:psf}) \citep[see][for details]{mclau08}. The best fitting model was derived by
calculating and minimizing $\chi^2$ as the sum of squared differences between model and observed intensities,
\begin{equation}
\chi^2=\sum_{i}{\frac{[I_{\rm obs}(R_i)-I_0\widetilde{I}_{\rm mod}^{*}(R_i|r_0)
       -I_{\rm bkg}]^2}{\sigma_{i}^{2}}},
\end{equation}
in which a background $I_{\rm bkg}$ was also fitted. The uncertainties of observed intensities listed in Table 3 were used as weights.

As an example, we plotted the fitting for one sample cluster in Figure 2. The observed intensity profile with extinction corrected is plotted as a function of logarithmic projected radius. The open squares are the data points included in the model fitting, while the crosses are points flagged as ``DEP'' or ``BAD'', which are not used to constrain the fit \citep{wang12a}. The best-fitting models, including the King model, Wilson model, and S\'{e}rsic model are shown with a solid line from the left to the right panel, with a fitted $I_{\rm bkg}$ added. The dashed lines represent the shapes of the PSFs for the filters used. There are some clusters showing individual isophotes with {\sc ellipse} intensities that showed irregular features or deviated strongly from their neighbors. As \citet{mclau08} reported, such bumps and dips may skew the following model fits. In these cases, we first derived the {\sc ellipse} output through a boxcar filter to make a smoothed cluster profile \citep{mclau08}, and then fitted these surface profiles using structural models. If some individual isophotes still cannot be well fitted, these points were deleted by hand, which were masked with ``BAD'' in Table 3.

\begin{figure}
\figurenum{2}\resizebox{\hsize}{!}{\rotatebox{0}{\includegraphics{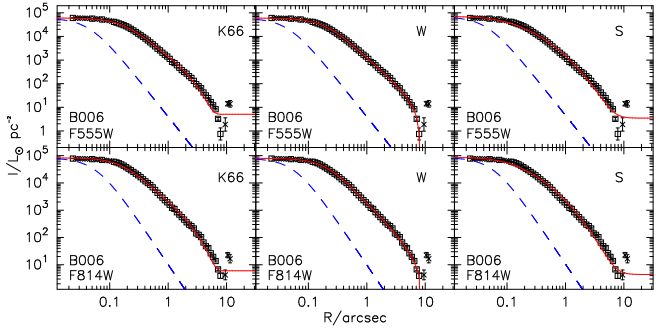}}}
\caption{Surface brightness profiles and model fits to one sample cluster B006, with the data of F555W and F814W band from top to bottom. The three panels in each line are fits to, from left to right: King model, Wilson model, and S\'{e}rsic model.}
\label{fig:fig2}
\end{figure}

Most profiles of the sample clusters were well fitted by the models, except for several clusters with different reasons. There are one or several bright objects at the intermediate radii of B097D, GC9, and M091; the shape is very loose for DAO38; the signal-to-noise ratio (SNR) is low for B205; there are one or several bright objects near the outer region of B328, B331, and BH11; the images are not well resolved for B092, B101, B145, BH05, and NB39; several clusters (B324, B330, B333, B468, GC3, GC7, BH04, and BH29) are lack of a central brightness concentration, which may be candidates of ``ring clusters''. The ``ring clusters'' have been reported in the Magellanic Clouds (MCs) and M33 \citep{mg03,hz06,wz11,roman12}, with irregular profiles such as bumps and dips which may not be attributed to the random fluctuations due to a few luminous stars. The images of these ``ring cluster'' candidates in our sample are displayed in Figure 3. There are three clusters (B018, B114, and B268) showing bumps only in the luminosity profiles of the F814W, with the intrinsic color of $(V-I)_0 =$ 0.95, 1.14, and 1.2, respectively. Some bright redder stars may locate at the intermediate radii of these clusters.

\begin{figure}
\figurenum{3} \resizebox{\hsize}{!}{\rotatebox{-0}{\includegraphics{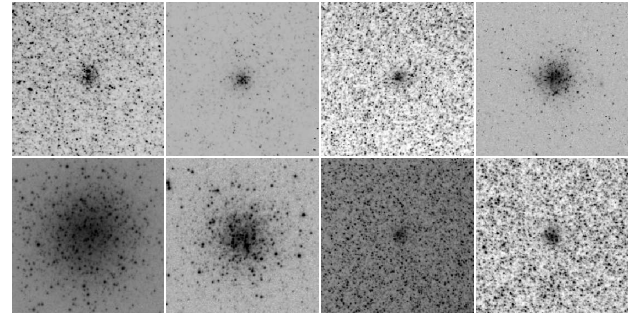}}}
\vspace{0.cm}
\caption{Images (F555W or F606W) of eight ``ring cluster'' candidates. From the upper left, these are B324, B330, B333, B468, GC3, GC7, BH04, and BH29.}
\label{fig:fig3}
\end{figure}

There are 15 clusters (B124, B127, B128, B148, B153, B167, B268, B331, BH05, BH12, M091, NB39, B064, B092, and B205) with high errors of $I_{\rm bkg}$. We checked the images carefully, and found that several reasons may account for those high errors: most of these clusters are not well resolved in the images; there are one or several bright objects in the intermediate and outer part of  clusters B331 and M091; clusters B064 and B205 only have images in one band (F300W), of which the SNRs are low. We should notice that if the fitted background is too high, the tidal radius may be estimated smaller artificially. Only one band of observations can be derived for six clusters (B009, B020D, B064, B092, B101, B205), and none of the bands is close to $V$ band. So, we would not include the six objects in the following discussions.

We used the same method mentioned in \citet{barmby09} and \citet{wang12a} to transform the magnitudes from ACS and WFPC2 to $V$ on the {\sc vegamag} scale \citep{holtzman95,siri05}. For clusters with available data of F555W or F606W band, we briefly used the extinction-corrected color $(V-I)_0$ or
$(B-V)_0$ to do the transformations, while for clusters
with no data of F555W or F606W band, we first transformed
the ACS or WFPC2 magnitudes to $I$ magnitude using the color $(B-I)_0$, and then computed the $V$ magnitude using the color $(V-I)_0$. The $BVI$ magnitudes and the reddening values are listed in Table 1. We estimated a precision of $\pm0.05$ mag in the transformation, which was propagated through the parameter estimates \citep{barmby07,wang12a}.

The mass-to-light ratios ($M/L$ values), which were used to derive the dynamical parameters, were determined from the population-synthesis models of \citet{bc03}, assuming a \citet{chab03} initial mass function. The ages and metallicities used to computed $M/L$ values in $V$ band
were derived as follows. An age of 13 Gyr with an uncertainty of $\pm2$ Gyr was adopted for old clusters, while the derived average ages (in Section \ref{brightness.sec}) were adopted for the young clusters.
The metallicities with uncertainties are listed in Table 1. The errors for $M/L$ of old clusters include
the uncertainties in age and metallicity, while
for young clusters, we simply adopted $10\%$ in $M/L$ as the errors as \citet{barmby09} did.

The basic structures and various derived dynamical parameters of the best-fitting models for each cluster are listed in Table 5 to Table 7, with a description of each parameter/column at the end of each table \citep[see for details of their calculation in][]{mclau08}. The uncertainties of these parameters were estimated by calculating their variations in each model that yields $\chi^2$ within 1 of the global minimum for a cluster, while combined in quadrature with the uncertainties in $M/L$ for the parameters related to it \citep[see][for details]{mm05}.

\subsection{Comparison with Previous Determinations}

In this paper, we determined structural parameters for 79 clusters in M31 by comparing their surface brightness profiles with three structural models. In Figure 4,
some estimated structural parameters for clusters in this paper were compared with those from previous studies
\citep[e.g.,][]{bhh02,barmby07,wang12a}.
\citet{bhh02,barmby07} presented structural and dynamical parameters in $V$ band for 51 clusters in our sample, while \citet{wang12a} determined structures for 10 GCs using King model. So we used the results on the bandpass close to $V$ band (e.g., F475W, F555W, and F606W) and fitted by King model for comparison. It is not unexpected to see that most of our parameters are larger than the results from \citet{bhh02,barmby07}, since the isophotes flagged as ``DEP'' in this paper may not be excluded from the
fitting process in \citet{bhh02}, resulting in
excessive weighting of the inner regions in the fits.
In addition, the ellipticities presented by \citet{bhh02} were averaged over different filters for each cluster,
while our results in Figure 4 were on the bandpass close to $V$ band. The cluster with the largest discrepancy of ellipticity is BH05, with an estimate of 0.19 by \citet{bhh02}, and 0.55 in this paper. As discussed above, the image of BH05 is not well resolved, so it is difficult to derive accurate ellipticities and structural parameters. \citet{barmby07} also concluded that the shapes of outer parts of faint clusters are strongly affected by the galaxy background or low SNRs, leading to a difficult business to accurately measure the ellipticities. The largest scatter of the $R_h$ is B018, which is $\sim5$ pc derived by \citet{barmby07}, while $\sim30$ pc in this paper. Although the PSFs and some calibration factors adopted here are slightly different from those of \citet{wang12a}, the parameters derived here are in good agreement with their results.

\begin{figure}
\figurenum{4}
\resizebox{\hsize}{!}{\rotatebox{0}{\includegraphics{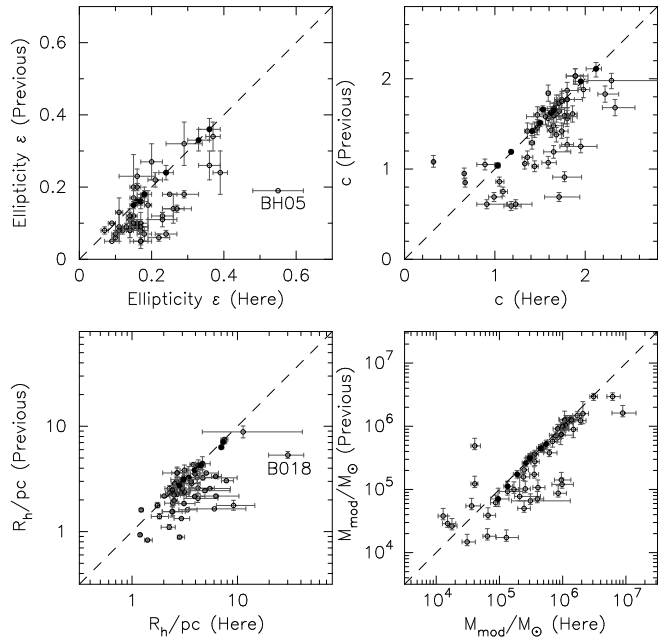}}}
\caption{Comparison of our newly obtained cluster structural parameters with previous measurements by \citet{bhh02,barmby07} (open circles) and \citet{wang12a}
(filled circles).}
\label{fig:fig4}
\end{figure}

\citet{str09,str11} presented observed velocity dispersions for a number of GCs in M31 using new high-resolution spectra from MMT/Hectochelle. These authors estimated the $M/L$ values in $V$ band for clusters using the virial masses and luminosities, since the virial masses are the nominal estimates of the ``global'' mass of
the system and are less sensitive to the accuracy of the King model fit \citep[see][for details]{str09,str11}.
Here we estimated the $M/L$ values in $V$ band from population-synthesis models by giving their metallicities and various ages \citep{barmby07,barmby09}.
In Figure 5, We presented the comparison of some parameters derived by \citet{barmby07,barmby09} and this paper with those from \citet{str11}. There is a large discrepancy between the $M/L$ values derived from the two methods, and most of the $M/L$ values from population-synthesis
models are larger than those from observed velocity dispersions. The model masses from \citet{barmby07,barmby09} and this paper are slightly larger than the virial masses from \citet{str11}, which were estimated using the half-mass radius and global velocity dispersion $\sigma_{\infty}$. The $\sigma_{p,\rm obs}$ values from \citet{str11} were the central velocity dispersions estimated using the observed velocity dispersions by integrating a known King model over the fiber aperture, and are consistent with the predicted line-of-sight velocity dispersions at the cluster center from \citet{barmby07,barmby09} and this paper.
However, there are few young clusters with velocity dispersion measured by \citet{str11}, so more observation data and analysis are needed to check the conclusion of the comparisons.

\begin{figure}
\figurenum{5}
\resizebox{\hsize}{!}{\rotatebox{0}{\includegraphics{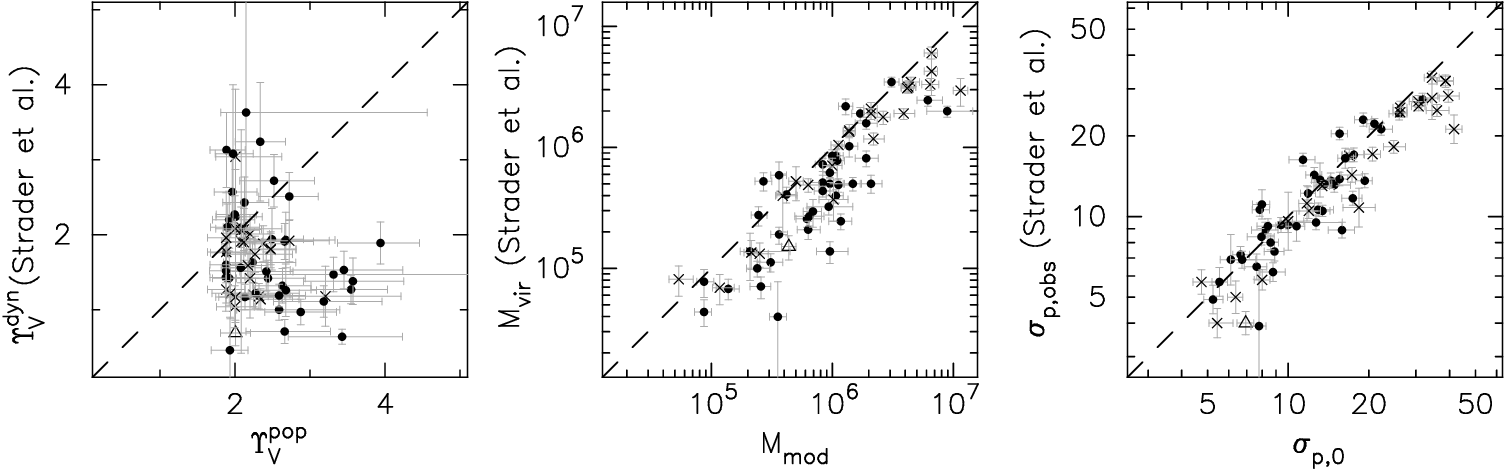}}}
\caption{Comparison of some dynamical parameters from \citet{barmby07} (crosses), \citet{barmby09} (triangles), and this paper (filled circles) with those from \citet{str11}.}
\label{fig:fig5}
\end{figure}

Figure 6 plots the correlations of velocity dispersion and mass with $M/L$ for M31 clusters. The left two panels show these parameters of clusters from \citet{str11}, while the right two panels show those derived by King model from \citet{barmby07,barmby09} and this paper. We can see that the $\sigma_{p,0}$ and $M_{\rm mod}$ from King-model fits show large dependence on the $M/L$. The old and young clusters show a distinct boundary of the $M/L$ values. There is no clear correlation for $M/L$ and $\sigma_{p,\rm obs}$ from \citet{str11}, while the decrease in $M/L$ toward lower masses is expected from dynamical evolution like evaporation, which is the steady loss of low-mass (high $M/L$) stars from the cluster driven by relaxation \citep{portegies10}. Considering the sensitive dependence on $M/L$ of $\sigma_{p,0}$, in the following discussion of the correlations of velocity dispersion with ellipticity and [Fe/H], we decided to use the $\sigma_{p,\rm obs}$ values from \citet{str11} for clusters in \citet{barmby07,barmby09} and this paper.

\begin{figure}
\figurenum{6}
\resizebox{\hsize}{!}{\rotatebox{0}{\includegraphics{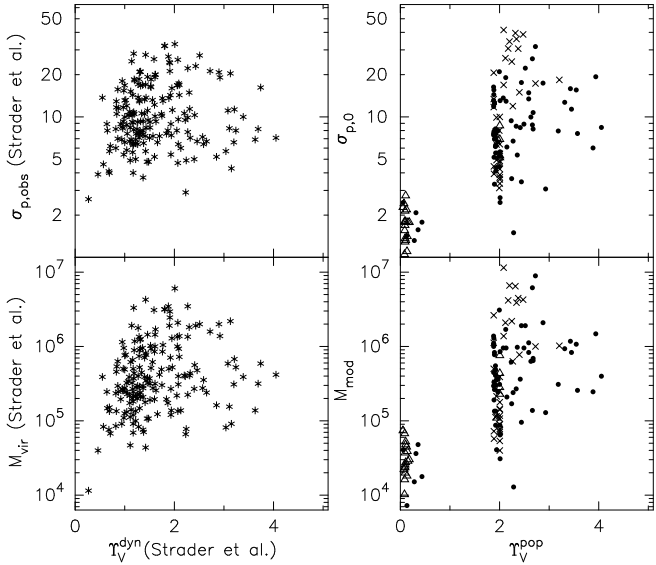}}}
\caption{Correlations of velocity dispersion and mass with $M/L$ in $V$ band. Left panels show clusters from \citet{str11} (asterisks) while right panels show clusters from \citet{barmby07} (crosses), \citet{barmby09} (triangles), and this paper (filled circles).}
\label{fig:fig6}
\end{figure}

\subsection{Comparison of Three Model Fittings}

In order to determine which model can describe the structure of clusters best, \citet{mm05} and \citet{mclau08} defined a relative $\chi^2$ index,
which compares the $\chi^2$ of the best fit of an ``alternate'' model with the $\chi^2$ of the best fit of King model,

\begin{equation}
\Delta=(\chi^2_{\rm alternate}-\chi^2_{\rm K66})/
(\chi^2_{\rm alternate}+\chi^2_{\rm K66}).
\end{equation}

It is evident that the ``alternate'' model is a better fit than King model if $\Delta$ is negative, while King model is a better fit if $\Delta$ is positive.

Figure 7 shows the relative quality of fit, $\Delta$ for
Wilson- and S\'{e}rsic-model fits versus King-model fits
for the sample clusters in this paper. The $\Delta$ values are plotted as a function of age and some structures, including the half-light radius $R_h$, the total model luminosity $L_{\rm mod}$, and the surface brightness over the half-light radius in the $V$ band $<\mu_V>_h$.
The circles refer to clusters with $R_{\rm last}\geq5R_h$,
while triangles present those with $R_{\rm last}<5R_h$,
where $R_{\rm last}$ is the most large radius for the available observation data. It can be seen that most clusters with $R_{\rm last}<5R_h$ are those having fainter luminosity $L_{\rm mod}$ or surface brightness $<\mu_V>_h$. \citet{mm05} presented that the fitting data out to $R_{\rm last}\geq5R_h$ can effectively constrain the model fitting to the outer regions, which are essential to determine which structural model best describes the clusters.
However, when $R_{\rm last}$ is small ($<5R_h$),
few data are available at large cluster radii, and
the fitting are mostly dependent on the inner part.
So, these clusters cannot be used to determine a preference of one model or the other. Similarly, we cannot conclude that these clusters are well fitted by both two models even $\Delta$ for them are small. \citet{mclau08} determined a catalogue of structural and dynamical parameters
for GCs in NGC 5128 using the three models, and
showed that the bright clusters ($L_{\rm mod}>10^5L_{\odot}$) are better fitted by Wilson model and S\'{e}rsic model, indicating that the halos of
clusters in NGC 5128 are more extended than what King model describes. There is no evident correlation of $\Delta$ with $L_{\rm mod}$ and $<\mu_V>_h$ for clusters in this paper.
Several studies \citep{elson87,roman12} showed that young clusters in the Large Magellanic Cloud (LMC) and M33 do not appear to be tidally truncated and seem to be better fitted by power-law profiles than King models, while old clusters show no clear differences between the qualities of the fittings. However, there is no correlation of $\Delta$ with age in Figure 7. We should notice that there are only six young clusters in our sample, and 13 clusters with no ages estimated in previous studies are assumed to be 13 Gyr. A large sample of young star clusters with precise age estimates are needed for the study on correlation of $\Delta$ with age. However, we do see that the King- and Wilson-model fits are better than the S\'{e}rsic-model fits. We concluded that clusters in M31 can be well fitted by both King model and Wilson model \citep[also reported in][]{barmby07,barmby09}.

\begin{figure}
\figurenum{7}
\resizebox{\hsize}{!}{\rotatebox{0}{\includegraphics{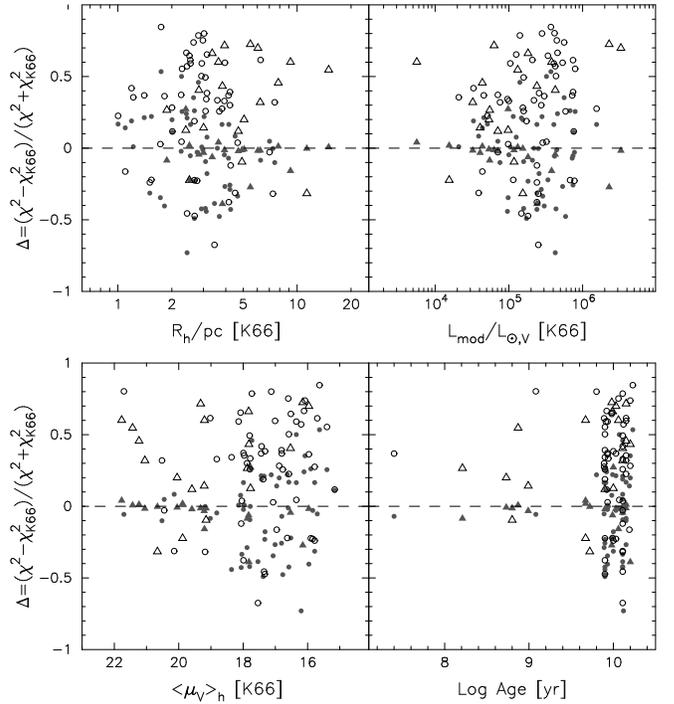}}}
\caption{Relative quality of fit for Wilson and S\'{e}rsic models (grey and open circles refer to the clusters with $R_{\rm last}\geq5R_h$, while grey and open triangles the clusters with $R_{\rm last}<5R_h$) versus King models for the sample clusters in this paper.}
\label{fig:fig7}
\end{figure}

Figure 8 compares the relative quality of fit,
$\Delta$ values with a number of structure parameters
($R_c$, $R_h$, $\mu_{V,0}$, $L_{\rm mod}$, $\sigma_{p,0}$, and $E_b$) for the sample clusters in this paper.
The grey and open circles show the physical properties of
clusters with $R_{\rm last}\geq5R_h$ derived from
Wilson- and S\'{e}rsic-model fits comparing to King-model fits, respectively. The triangles refer to clusters with $R_{\rm last}<5R_h$. There are some clusters with comparable $\chi^2$, but large discrepancy of $R_h$ and $L_{\rm mod}$ values for King- and Wilson-model fits.
We can see that most of these clusters have $R_{\rm last}<5R_h$. As discussed above, the few constrain of the fitting to outer regions results in much different extrapolations of models, and it is hard to determine
which model does the correct fitting in those cases \citep{mm05}. Most parameters derived from Wilson model are slightly larger than those from King model, while parameters derived by S\'{e}rsic model are smaller than those from King model, especially for $R_c$, $\sigma_{p,0}$, and $E_b$. \citet{barmby07} concluded that rather than an intrinsic difference between clusters in M31 and other galaxies, the preference for King models over Wilson models for M31 clusters is due to some subtle features of the observations.

\begin{figure}
\figurenum{8}
\resizebox{\hsize}{!}{\rotatebox{0}{\includegraphics{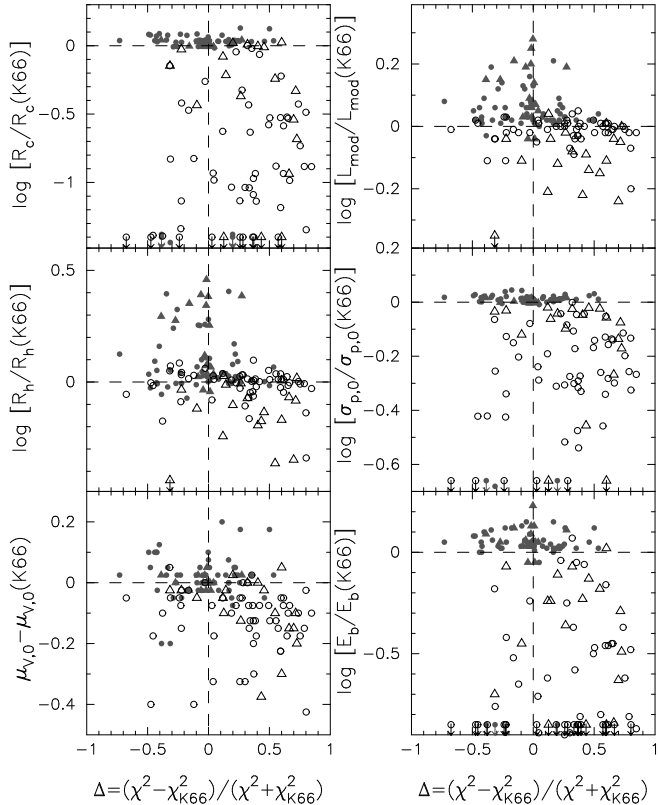}}}
\caption{Comparison of some parameters for Wilson and S\'{e}rsic models versus King models for the sample clusters in this paper, including the projected core radius $R_c$, the projected half-light radius $R_h$,
the central surface brightness in the $V$ band $\mu_{V,0}$, the total model luminosity $L_{\rm mod}$,
the predicted central line-of-sight velocity dispersion $\sigma_{p,0}$, and the global binding energy $E_b$. Symbols are as in Fig. 7.}
\label{fig:fig8}
\end{figure}

\section{DISCUSSION}
\label{discussion.sec}

We combined the newly derived parameters here with
those derived by King-model fits for M31 young massive clusters (YMCs) \citep{barmby09}, M31 globulars \citep{barmby07}, M31 extended clusters (ECs) \citep{huxor11}, and MW globulars \citep{mm05}
to construct a large sample to look into the correlations
between the parameters. The ellipticities and galactocentric distances for the MW GCs are
from \citet{harris96} (2010 edition). The parameters used in the following discussion for M31 GCs \citep{barmby07}
are those derived on the bandpass close to $V$ band
(e.g., F555W and F606W), while the data from
bluer filters are preferred for the YMCs in \citet{barmby09}. The metallicities for most young clusters in \citet{barmby09} were assumed to be solar metallicity, while metallicities from \citet{perina10} were adopted for
some older clusters (B083, B222, B347, B374, and NB16).
\citet{huxor11} derived structure parameters of 13 ECs, including the core radius, half-light radius, tidal radius, and the central surface brightness in magnitude.
The metallicities of 4 ECs (HEC4, HEC5, HEC7 and HEC12)
were determined by \citet{mackey06}, and the integrated cluster masses for them were derived here using the absolute magnitudes and $M/L$ values in $V$ band from population-synthesis models \citep[see][for the details]{wang12a}.

\subsection{Galactocentric Distance}

Figure 9 shows structural parameters as a function of galactocentric distance $R_{\rm gc}$ for the new large sample clusters in M31 and the MW. Some global trends can be seen. Both $R_h$ and $r_t$ increase with the $R_{\rm gc}$ as expected, although the trend is largely driven by the ECs from \citet{huxor11}. The star clusters with large $R_{\rm gc}$ can keep more stars with weaker tides relatively to those located at bulge and disk. \citet{Georgiev09} explained that the $R_h$ and $r_t$ of the clusters in the halo of galaxies, which may be accreted into the galaxies from dwarf galaxies, can expand due to the change from strong to a weaker tidal field. \citet{bhh02} concluded that the correlation of $R_h$ with $R_{\rm gc}$ reflects physical formation conditions as suggested by \citet{vanden91} for GCs in the MW. However, \citet{str12} found that no clear correlation
between $R_h$ and $R_{\rm gc}$ exists beyond $R_{\rm gc}\sim15$ kpc for GCs in NGC 4649, and they suggested that the sizes of GCs are not generically set by tidal limitation. It can be seen that there are no compact clusters at large radii ($R_{\rm gc}>40$ kpc) in the MW, while in M31, there is a bimodality in the size distribution of GCs at large radii \citep[also reported in][]{huxor11,wang12a}. It is interesting that there are few GCs having $R_h$ in the range from 8 to 15 pc at large radii ($R_{\rm gc}>40$ kpc) in M31. There are three clusters (B124, B127, and NB39) with large $R_h$ and $r_t$ at small $R_{\rm gc}$, leading to more diffuse trends. However, the total model masses for them are $10^{6.95}$, $10^{6.79}$, and $10^{5.93} M_{\odot}$, respectively, meaning that they are massive clusters. It is not unexpected that these massive clusters can contain more stars than other clusters, although they are at small $R_{\rm gc}$. The luminosities of clusters decrease with increasing $R_{\rm gc}$, implying that either strong dynamical friction drives predominantly more massive GCs inwards, or massive GCs may favor to form in the nuclear regions of galaxies with the higher pressure and density \citep{Georgiev09}. However, the lack of faint clusters with small galactocentric distance may be due to selection effects, since these clusters are difficult to detect against the bright background near M31 center, which is also reported by \citet{barmby07} using the correlation of central surface brightness with galactocentric distance. The metallicities of clusters decrease with increasing $R_{\rm gc}$, indicating that metal-rich clusters are typically located at smaller galactocentric radii than metal-poor ones, although with large scatter. \citet{vanden91} found that metal-rich clusters ([Fe/H]$\geq-0.8$) in M31 seem to form in a rotating disk extending to $R_{\rm gc}\simeq5$ kpc. The metal-rich GCs may have undergone accelerated internal evolution due to strong tidal shocks \citep{str11} from both bulge and disk passages. There are two matal-poor clusters (B114, B324) at small $R_{\rm gc}$, which are at the bottom-left in the [Fe/H]-$R_{\rm gc}$ panel. The diffuse distribution of these parameters vs. $R_{\rm gc}$ may be caused by the different galactocentric distance used, which is true three-dimensional distances for Galactic GCs while projected radii for M31 clusters.

\begin{figure}
\figurenum{9}
\resizebox{\hsize}{!}{\rotatebox{0}{\includegraphics{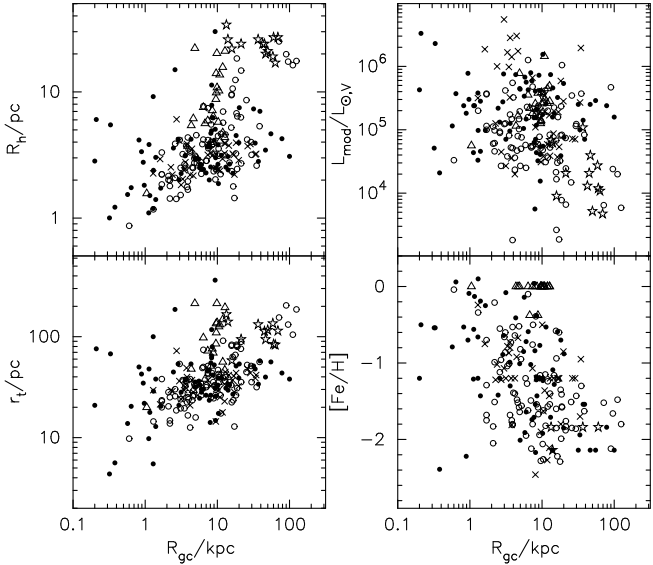}}}
\caption{Structural parameters vs. galactocentric distance $R_{\rm gc}$. The filled circles are the sample clusters in M31 (this paper), the open circles are Galactic GCs \citep{mm05}, the crosses are M31 GCs \citep{barmby07},
the open triangles are M31 YMCs \citep{barmby09},
and the open stars are M31 ECs \citep{huxor11}.}
\label{fig:fig9}
\end{figure}

\subsection{Ellipticity}

\begin{figure}
\figurenum{10} \resizebox{\hsize}{!}{\rotatebox{-90}{\includegraphics{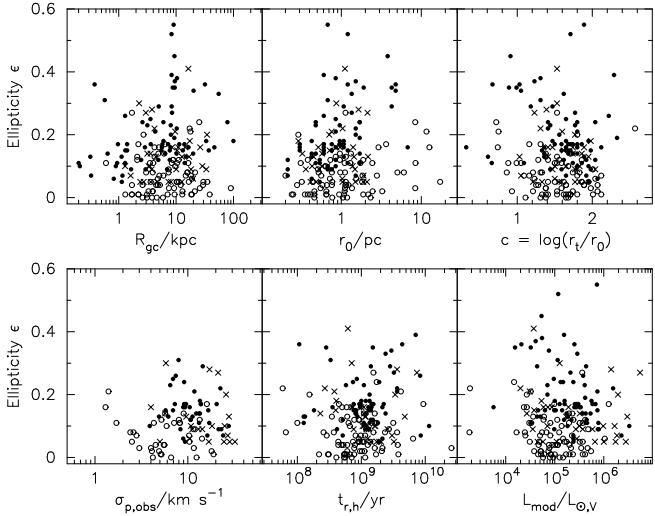}}}
\caption{Ellipticity as a function of galactocentric distance, metallicity, the observed velocity dispersion, and structure parameters. Symbols are as in Fig. 9.}
\label{fig:fig10}
\end{figure}

Figure 10 shows the distribution of ellipticity with galactocentric distance, metallicity, and some structure parameters for clusters in the MW and M31, which may show clues to the primary factor for the elongation of clusters: rotation and velocity anisotropy, cluster mergers, ``remnant elongation'', and galaxy tides \citep{larsen01,barmby07}. \citet{Geyer83} presented that the tidal forces can only distort a cluster's outer region, while the elongation of the inner parts are due to its own gravitational potential and the total orbital angular momentum.

1) Rotation and velocity anisotropy. Cluster rotation is the generally accepted explanation for cluster flattening \citep{dp90}. Dynamical models show that internal relaxation coupled to the external tides will drive a cluster toward a rounder shape over several relaxation times \citep[see][and references therein]{harris02}.
\citet{bhh02} presented that more compact clusters which experience more relaxation processes, and clusters with larger velocity dispersions which rotate more slowly, should be rounder.
\citet{barmby07} presented that dynamical evolution could reduce the initial flattening caused by rotation or velocity anisotropy, indicating that more evolved clusters would be rounder. In order to check these predictions, we show the correlations of $r_0$, $c$, the observed velocity dispersion $\sigma_{p,\rm obs}$, and $t_{r,h}$ with the ellipticity. The observed velocity dispersion $\sigma_{p,\rm obs}$ for M31 clusters are from \citet{str11}, while $\sigma_{p,\rm obs}$ for clusters in the MW are from \citet{mm05}. All the $\sigma_{p,\rm obs}$ have been extrapolated to their central values with an aperture of 0. There is no clear correlation of ellipticity with these parameters. However, we do see that more compact clusters--smaller $r_0$ and larger $c$--are more rounder, which is consistent with previous studies \citep{bhh02}.

2) Cluster mergers and ``remnant elongation''. \citet{vanden84} presented that ellipticity correlates strongly with luminosity for clusters in LMC: more luminous clusters are more flattened, while \citet{vanden96} presented that the most flattened GCs in both the MW and M31 are also brightest. This may be explained by the cluster mergers and ``remnant elongation'', which is from some clusters' former lives as dwarf galaxy nuclei. So, the larger ellipticities for clusters located at large galactocentric radii of M31 may be related to the merger or accretion history that M31 has experienced \citep[e.g.][]{bekki10,mackey10,huxor11}. However, no clear correlation of ellipticity with luminosity can be seen.

3) Galaxy tides. A strong tidal field might rapidly destroy the velocity anisotropies, and force an initially triaxial, rapidly rotating elliptical GC to a more isotropic distribution and spherical shape, while weak tidal fields are unable to change the initial shapes of GCs \citep{goodwin97}. It seems plausible that clusters located at different galactocentric radii or different galaxies have various distributions of ellipticities, due to the diverse galaxy tides. \citet{harris02} and \citet{barmby07} found the distributions of ellipticities for M31 and NGC 5128 are very similar, but differ from the MW distribution, which has few very round clusters. No clear correlation of ellipticity with $R_{\rm gc}$ can be seen for clusters in these galaxies \citep[also see][]{barmby07}. However, the innermost clusters are slightly more spherical \citep{bhh02}, which may due to the strong tidal field near galaxy center that reduces ellipticities of them. Some clusters located at large projected radii of M31 do have larger ellipticities than most GCs in M31 and the MW, which may be caused by galaxy tides coming from satellite galaxies \citep{wang12a}.

\citet{roman12} summarized the orientation of a large sample of clusters in M33 and found that the distribution of PAs shows a strong peak at $-55^{\circ}$, which is close to the direction towards M31. These authors suggested that the elongation of clusters in M33 may be attributed to the tidal forces of M31, considering that a recent encounter between M33 and M31 \citep{McConnachie09,putman09,roman10,bernard12} may have led to significant effects on the properties of M33 disk. Figure 11 depicts the distribution of PAs of star clusters in this paper and 34 clusters from \citet{barmby07}, which were not included in \citet{bhh02}. No clear trend is present in the orientation vectors towards M33, although a small peak at $-60^{\circ}$ does exist in the distribution of the PAs. Here we concluded that the elongation of clusters seems to be due to various factors \citep{barmby07}.

\begin{figure}
\figurenum{11} \resizebox{\hsize}{!}{\rotatebox{0}{\includegraphics{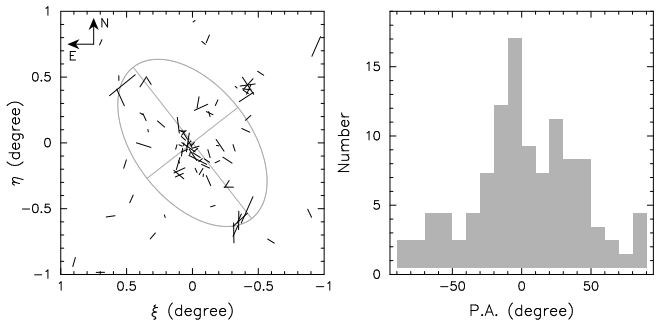}}}
\caption{($Left~panel$) Cluster elongations and orientations shown with respect to M31. The vector sizes are correlated with the ellipticities. The ellipse has a radius of 10 kpc and is flattened to $b/a = 0.6$. ($Right~panel$) Distribution of the P.A.s of star clusters in M31.}
\label{fig:fig11}
\end{figure}

\subsection{Metallicity}

Figure 12 plots structural parameters as a function of [Fe/H] for the new large sample clusters in M31 and the MW. The trends of the parameters with [Fe/H] nearly disappear
when we add the data for the M31 YMCs, which were fitted with solar metallicities \citep{barmby09}. However, we noticed that the metallicities for most of these YMCs obtained by \citet{kang12} are poorer than [Fe/H]$=-1.1$.
If we do not consider the YMCs in \citet{barmby09}, it is evident that the metal-rich clusters have smaller average values of $R_h$ than those of metal-poor ones \citep{larsen01,bhh02}. \citet{str12} presented that the sizes of metal-rich GCs are smaller than the metal-poor ones in NGC 4649, which is a massive elliptical
galaxy in the Virgo galaxy cluster. These authors explained that as an intrinsic size difference rather than projection effects. \citet{sippel12} carried out N-body simulations of metallicity effects on cluster evolution, and found that there is no evident difference for the half-mass radii of metal-rich and metal-poor cluster models. So, they explained that metal-rich and metal-poor clusters have similar structures, while metallicity effects combined with dynamical effects such as mass segregation produce a larger difference of the half-light radii. \citet{barmby07} also found that $R_h$ decreases with increased metallicity for GCs in the MW, the MCs, Fornax dwarf spheroidal, and M31, except for GCs in NGC 5128. An evident feature is that these four ECs are all metal-poor and have large $R_h$. No clear correlation of model luminosity $L_{\rm mod}$ with [Fe/H] is present, but we do see that the metal-rich clusters tend to be more luminous. Both the observed velocity dispersion $\sigma_{p,\rm obs}$
and central ``escape'' velocity $\nu_{\rm esc,0}$ increase with metallicity. \citet{Georgiev09} presented that the higher $\nu_{\rm esc}$ of more metal-rich clusters in the observation may reflect the metallicity dependence of
the terminal velocities of the stellar winds, since the $\nu_{\rm esc}$ of a metal-rich cluster should be higher to retain the fast stellar winds.

\begin{figure}
\figurenum{12} \resizebox{\hsize}{!}{\rotatebox{0}{\includegraphics{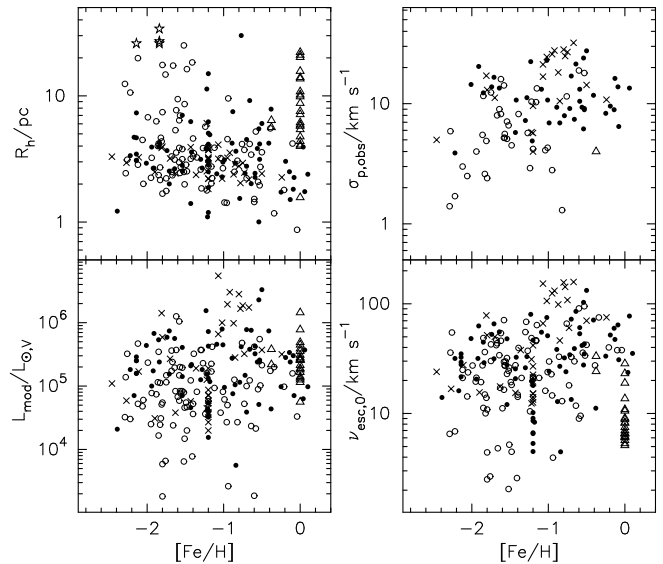}}}
\caption{Structural parameters and the observed velocity dispersion as a function of [Fe/H]. Symbols are as in Fig. 9.}
\label{fig:fig12}
\end{figure}

\subsection{Core-collapsed Clusters}

Core-collapsed clusters in general show a power-law slope in the central surface brightness profiles, which can be better fitted by a power-law model than King model \citep{bhh02}. \citet{ng06} presented that the process of core collapse can be divided into two stages. In the first stage, stars are driven to the halo of the cluster due to close encounters. Stellar evaporation occurs and the core contracts due to energy conservation. In the second stage, the low-mass stars are scattered to high velocities
and escape to the halo, while the high-mass stars sink to the core due to energy equipartition. The increasing core density also increases the interaction rate of the binaries and single stars, which can generate energy in the
core, reverse the contraction process, and produce an expansion. After a long time, the core shrinks again, and the process repeats. M15, which has a central cusp, may be at an intermediate state between the extremes of collapse and expansion \citep{dull97}. \citet{chatt12} presented that the core-collapsed clusters are those that have reached or are about to reach the ``binary burning'' stage, while the non core-collapsed clusters are still contracting under two-body relaxation. \citet{tkd95} presented a catalogue of surface brightness profiles of 125 Galactic GCs, and classified 16\% of their sample as core-collapsed clusters and 6\% as core-collapsed candidates. \citet{mclau08} noticed that a number of clusters (more than half of their sample) in NGC 5128 have the index $n>2$ derived from S\'{e}rsic model, indicating that these clusters have strongly peaked central density profiles. These authors explained that the PSF may flatten the models, and then the cuspy central density profiles are shown relatively. \citet{ng06} presented that Galactic clusters with steep cusps are all close to the center of the Galaxy, indicating that an increased incidence of tidal shocking \citep{glo99} may accelerate the core collapse process. \citet{bhh02} also concluded that most of the M31 core-collapsed candidates are within 2 kpc of
the center of M31.

Figure 13 shows the distributions of galactocentric distance, and some structure parameters derived from S\'{e}rsic-model fitting for clusters in this paper and M31 GCs in \citet{barmby07}. There are 25 clusters bested fitted by S\'{e}rsic model ($\chi^2_{\rm S}<\chi^2_{\rm K66}$ and $\chi^2_{\rm S}<\chi^2_{\rm W}$), and 11 of them have $n>2$, indicating that they have cuspy central density profiles. We assumed these 11 clusters to be possible core-collapsed clusters. In fact, the clusters with $n>2$ do have smaller $R_c$ and larger $R_h/R_c$ than their counterparts with $n<2$. \citet{bhh02} presented several clusters (B011, B064, B092, B123, B145, B231, B268, B343, and BH18) as core-collapsed candidates in M31. However, most of these clusters are best fitted by King model or Wilson model in this paper except B145 and B343. We checked the images of these clusters and found that
the images of B123, B231, and B268 are not fully resolved, while the SNRs of images for B064 and B092 are low. Actually, the discrimination between the core-collapsed clusters and the ``King model clusters'' \citep{mh97} may often become unclear for several reasons: (1) statistical noise due to some unresolved bright stars in the cores of clusters; (2) the similarity between the high-concentration King-model and power-law profiles \citep[see][for details]{mh97}. B145 is best fitted by S\'{e}rsic model in this paper, but $n=1.5$. The cluster B343 shows a cuspy density profile in the center, and has been classified as a core-collapsed cluster \citep{bendinelli93,grill96}. We also find a better fitting by S\'{e}rsic model for it with $n=3.35$. The distribution of core-collapsed cluster candidates is not constrained to the center of M31. There are only two out of the 11 clusters with $R_{\rm gc} < 2$ kpc, while four ones with $R_{\rm gc} > 15$ kpc.

\begin{figure}
\figurenum{13} \resizebox{\hsize}{!}{\rotatebox{0}{\includegraphics{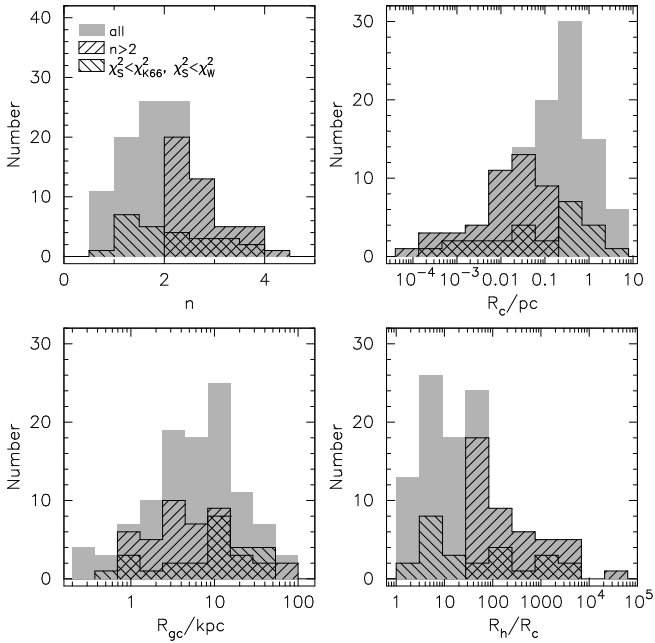}}}
\caption{The distributions of the galactocentric distance and several structure parameters derived from S\'{e}rsic-model fitting for clusters in this paper and M31 GCs in \citet{barmby07}.}
\label{fig:fig13}
\end{figure}

\subsection{The Fundamental Plane}

\begin{figure}
\figurenum{14} \resizebox{\hsize}{!}{\rotatebox{0}{\includegraphics{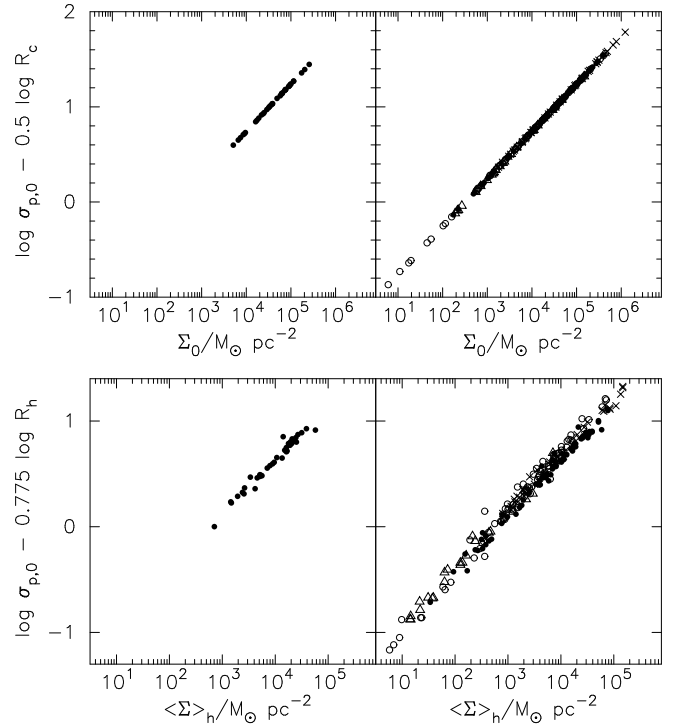}}}
\caption{Evidence of an FP of the cluster parameters,
which is defined in terms of central velocity dispersion $\sigma_{p,0}$, radius $R_c$ or $R_h$, and surface mass density $<\Sigma>_0$ or $<\Sigma>_h$. Symbols are as in Fig. 9.}
\label{fig:fig14}
\end{figure}

It has been widely noticed that GCs do not occupy the full
four-dimensional parameter space (concentration $c$, scale radius $r_0$, central surface brightness $\mu_{V,0}$,
and central $M/L$ or velocity dispersion $\sigma_0$) instead locate in a remarkably narrow ``fundamental plane'' (FP). It is interesting to learn which of the structural and dynamical properties are either universal or dependent on characteristics of their parent galaxies \citep{mh97}.
\citet{dm94} explored some tight correlations between various properties using a large sample of Galactic clusters, and \citet{djor95} defined an FP for GCs in terms of velocity dispersion, radius, and surface brightness.
\citet{saito79} investigated the mass-binding energy relationship for a few GCs and ellipticals, while \citet{mclau00} defined an FP using the binding energy and luminosity, which is formally different from that of \citet{djor95}. \citet{harris02} found that the NGC 5128 GCs describe a relation between binding energy and luminosity tighter than in the MW. Here we show the two forms of the FP for the new large sample clusters in the MW and M31.

Figure 14 shows the mass-density-based FP relations.
The left two panels show the correlations of the properties derived with the $M/L$ values from the velocity dispersions \citep{str11}, using King model and the same fitting process in this paper (Section \ref{fit.sec}),
while the right two panels show the correlations of the properties derived with the $M/L$ values from population-synthesis models. \citet{barmby09} presented the surface-brightness-based FP relations and found a large offset between the young M31 clusters and old clusters. They explained that as a result from lower $M/L$ values for the young clusters in M31. It is obvious that the velocity dispersion, characteristic radii, and surface density for these clusters show tight relations, both on the core and half-light scales. The exist of FP for clusters
strongly reflects some universal physical conditions and processes of cluster formation.

Figure 15 shows the correlation of binding energy with the
total model mass. The left panel shows the $M_{\rm mod}$ and $E_b$ derived with the $M/L$ from the directly observed velocity dispersions \citep{str11}, using King model and the same fitting process in this paper (Section \ref{fit.sec}), while the right panel shows those properties derived with the $M/L$ from population-synthesis models. All the clusters locate in a remarkably tight region although in the widely different galaxy environments, which is consistent with the previous studies \citep{barmby07,barmby09}. \citet{barmby07} concluded that the scatter around this relation is so small that the structures of star clusters may be far
simpler than those scenarios derived from theoretical arguments.

\begin{figure}
\figurenum{15} \resizebox{\hsize}{!}{\rotatebox{0}{\includegraphics{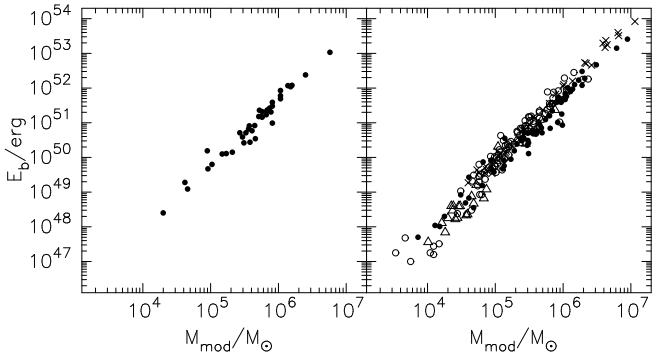}}}
\caption{Evidence of an FP of the cluster parameters,
which is defined in terms of binding energy $E_b$ and the
total model mass $M_{\rm mod}$. Symbols are as in Fig. 9.}
\label{fig:fig15}
\end{figure}

\section{SUMMARY}

High-resolution imaging can be derived from {\it HST} observations for M31 star clusters. In this paper, we presented surface brightness profiles for 79 clusters,
which were selected from \citet{bhh02} and \citet{mackey07}. Structural and dynamical parameters were derived by fitting the profiles to three different models, including King model, Wilson model, and S\'{e}rsic model. We found that in the majority of cases, King models fit the M31 clusters as well as Wilson models, and better than S\'{e}rsic models.

We discussed the properties of the sample GCs here
combined with GCs in the MW \citep{mm05} and clusters in M31 \citep{barmby07,barmby09,huxor11}. In general, the properties of the M31 and the Galactic clusters
fall in the same regions of parameter spaces.
There is a bimodality in the size distribution of M31 clusters at large radii, which is different from their Galactic counterparts. There are 11 clusters in M31 best fitted by S\'{e}rsic models with index $n>2$, meaning that they have cuspy central density profiles, which are classified as core-collapsed cluster candidates. We investigated two forms of the FP, including the correlation of velocity dispersion, radius, and surface density, and the correlation of binding energy with the total model mass. The tight correlations of cluster properties indicate a tight FP for clusters, regardless of their host environments, which is consistent with previous studies \citep{barmby07,barmby09}. In addition, the tightness of the relations for the internal properties indicates some physical conditions and processes of cluster formation in different galaxies.

\acknowledgments We thank the anonymous referee for providing a rapid and thoughtful report that helped improve the original manuscript greatly. We would like to thank Dr. McLaughlin for his help in deriving the parameters of the three structure models, and Richard Hook in understanding the TinyTim package, and Pauline Barmby in understanding the PSF of {\it HST}, and Christopher Willmer in providing the updated table of absolute solar magnitudes. This work was supported by the Chinese National Natural Science Foundation grands No. 10873016, and 10633020, and by National Basic
Research Program of China (973 Program), No. 2007CB815403.

\clearpage

\begin{sidewaystable}
\begin{center}
\small
\caption{Integrated measurements for 79 sample clusters in M31.} \vspace{2mm} \label{t1.tab}
\begin{tabular}{cccccccccccc}
\tableline\tableline
Name & $\epsilon^a$ & $\theta^a$ & $\epsilon^b$ & $\theta^b$ & $B$ & $V$ & $I$ & $ R_{\rm gc} $ & $E(B-V)$ & ${\rm [Fe/H]}$ & ${\rm Age}$\\
     &    & (deg E of N)  &     & (deg E of N) & (Vegamag) & (Vegamag) & (Vegamag) & (kpc) &   &  & (Gyr)\\
(1) &  (2)  &  (3) &  (4) & (5) &  (6)  & (7) &  (8)  &  (9) & (10) &  (11) &(12) \\
\hline
 B006     & $0.14     \pm 0.01    $ & $-40      \pm 8       $ & $0.24     \pm 0.02    $ & $-45     \pm8       $ & 16.49    & 15.50    & 14.33    & 6.43     & 0.11     & $-0.59    \pm 0.41    $ & 13.25    \\
 B011     & $0.12     \pm 0.01    $ & $21       \pm 8       $ & $0.15     \pm 0.01    $ & $32      \pm7       $ & 17.46    & 16.58    & 15.56    & 7.73     & 0.09     & $-1.71    \pm 0.24    $ & 15.85    \\
 B012     & $0.17     \pm 0.02    $ & $24       \pm 8       $ & $0.14     \pm 0.01    $ & $18      \pm7       $ & 15.86    & 15.09    & 14.03    & 5.78     & 0.11     & $-1.91    \pm 0.21    $ & 8.00     \\
 B018     & $0.19     \pm 0.01    $ & $-9       \pm 5       $ & $0.15     \pm 0.02    $ & $-5      \pm8       $ & 18.37    & 17.53    & 16.38    & 9.31     & 0.20     & $-0.77    \pm 0.39    $ & 1.20     \\
 B027     & $0.18     \pm 0.02    $ & $11       \pm 7       $ & $0.16     \pm 0.03    $ & $-12     \pm9       $ & 16.49    & 15.56    & 14.41    & 6.02     & 0.19     & $-1.64    \pm 0.16    $ & 15.75    \\
 B030     & $0.16     \pm 0.01    $ & $-12      \pm 12      $ & $0.13     \pm 0.01    $ & $-37     \pm6       $ & 18.75    & 17.39    & 15.68    & 5.66     & 0.57     & $-0.14    \pm 0.26    $ & 9.62     \\
 B045     & $0.12     \pm 0.01    $ & $-13      \pm 6       $ & $0.16     \pm 0.02    $ & $-13     \pm7       $ & 16.72    & 15.78    & 14.51    & 4.90     & 0.18     & $-1.01    \pm 0.50    $ & 11.40    \\
 B058     & $0.09     \pm 0.01    $ & $-27      \pm 8       $ & $0.11     \pm 0.01    $ & $-36     \pm9       $ & 15.81    & 14.97    & 13.87    & 6.96     & 0.12     & $-1.02    \pm 0.21    $ & 8.01     \\
 B068     & $0.21     \pm 0.02    $ & $31       \pm 3       $ & $0.18     \pm 0.02    $ & $37      \pm4       $ & 17.60    & 16.39    & 14.84    & 4.32     & 0.42     & $-0.41    \pm 0.17    $ & 7.90     \\
 B070     & $0.12     \pm 0.01    $ & $1        \pm 10      $ & $0.16     \pm 0.02    $ & $-24     \pm11      $ & 17.61    & 16.76    & 15.72    & 2.46     & 0.12     & $-1.42    \pm 0.43    $ & 8.75     \\
\tableline
\end{tabular}
\end{center}
\footnotetext[1]{$\epsilon$ and $\theta$ of bluer filters.}
\footnotetext[2]{$\epsilon$ and $\theta$ of redder filters.}
\end{sidewaystable}

\begin{table}
\centering
\caption{Calibration Data for {\sl HST} images.} \vspace{0mm} \label{t2.tab}
\begin{tabular}{ccccccc}
\tableline\tableline
Filter   & Pivot $\lambda$ & $R_{\lambda}^a$ & Zeropoint$^b$ & $M_{\odot}^c$ & Conversion Factor$^d$ &  Coefficient$^e$\\
  & (\AA) &  &  &  &  & \\
(1) &  (2)  &  (3) &  (4) & (5) &  (6)  & (7) \\
\hline
\multicolumn{7}{c}{Calibration Data for ACS/WFC images} \\
\hline
F435W  &  4318.9 & 4.20   & 25.779  &  5.459     &  3.1693     &   27.031     \\
F475W  &  4746.9 & 3.72   & 26.168  &  5.167     &  1.6926     &   26.739     \\
F555W  &  5361.0 & 3.19   & 25.724  &  4.820     &  1.8508     &   26.392     \\
F606W  &  5921.1 & 2.85   & 26.398  &  4.611     &  0.8207     &   26.183     \\
F814W  &  8057.0 & 1.83   & 25.501  &  4.066     &  1.1349     &   25.638     \\
\hline
\multicolumn{7}{c}{Calibration Data for WFPC2 images} \\
\hline
F300W  &  2986.8 & 5.66   & 21.448  &  6.061     &   297.9614  &   27.633     \\
F450W  &  4555.4 & 3.93   & 24.046  &  5.263     &   13.0545   &   26.835    \\
F555W  &  5439.0 & 3.14   & 24.596  &  4.804     &   5.1542    &   26.376   \\
F606W  &  5996.8 & 2.81   & 24.957  &  4.581     &   3.0100    &   26.153   \\
F814W  &  8012.2 & 1.85   & 23.677  &  4.074     &   6.1342    &   25.646   \\
\tableline
\end{tabular}
\footnotetext[1]{$A_{\lambda}=R_{\lambda} \times E(B-V)$.}
\footnotetext[2]{Additive conversion between surface brightness in counts s$^{-1}$ arcsec$^{-2}$ and magnitude in mag arcsec$^{-2}$.}
\footnotetext[3]{Updated absolute magnitude of the sun (C. Willmer, private communication).}
\footnotetext[4]{Multiplicative conversion between surface brightness in counts s$^{-1}$ arcsec$^{-2}$ and intensity in $L_{\odot}$ pc$^{-2}$.}
\footnotetext[5]{Additive conversion between surface brightness in magnitude in mag arcsec$^{-2}$ and intensity in $L_{\odot}$ pc$^{-2}$.}
\end{table}

\begin{table}
\centering
\caption{The intensity profiles for 79 sample clusters in M31.} \vspace{0mm} \label{t3.tab}
\begin{tabular}{cccccccc}
\tableline\tableline
Name & Detector & Filter & $R$ & $I$ & Uncertainty & Flag \\
  &  &  & (arcsec) &  ($L_{\odot}$~pc$^{-2}$) & ($L_{\odot}$~pc$^{-2}$) & \\
(1) &  (2)  &  (3) &  (4) & (5) &  (6)  & (7) \\
\hline
 B006     & WFPC2/PC   & F555W    & 0.0234     & 44001.773    & 155.603      & OK    \\
          & WFPC2/PC   & F555W    & 0.0258     & 43776.562    & 171.971      & DEP   \\
          & WFPC2/PC   & F555W    & 0.0284     & 43528.629    & 188.743      & DEP   \\
          & WFPC2/PC   & F555W    & 0.0312     & 43258.270    & 206.935      & DEP   \\
          & WFPC2/PC   & F555W    & 0.0343     & 42964.875    & 227.137      & DEP   \\
          & WFPC2/PC   & F555W    & 0.0378     & 42647.070    & 249.659      & DEP   \\
          & WFPC2/PC   & F555W    & 0.0415     & 42303.609    & 274.847      & DEP   \\
          & WFPC2/PC   & F555W    & 0.0457     & 41919.195    & 299.011      & DEP   \\
          & WFPC2/PC   & F555W    & 0.0503     & 41381.914    & 308.833      & OK    \\
          & WFPC2/PC   & F555W    & 0.0553     & 40666.176    & 309.009      & DEP   \\
          & WFPC2/PC   & F555W    & 0.0608     & 39884.902    & 322.641      & DEP   \\
          & WFPC2/PC   & F555W    & 0.0669     & 38966.234    & 328.825      & DEP   \\
          & WFPC2/PC   & F555W    & 0.0736     & 37925.949    & 316.808      & OK    \\
\tableline
\end{tabular}
\end{table}

\begin{table}
\centering
\caption{Coefficients for the PSF models.} \vspace{2mm} \label{t4.tab}
\begin{tabular}{lccccc}
\tableline\tableline
Detector  &Filter  &  $r_0$    & $\alpha$ & $\beta$   \\
          &        & (arcsec)  &          &           \\
  (1)     &  (2)   &  (3)      &  (4)     &  (5)      \\
\hline
ACS/WFC   &  F435W &  0.068    &   3    &   3.80     \\
          &  F475W &  0.064    &   3    &   3.60     \\
          &  F555W &  0.057    &   3    &   3.39     \\
          &  F606W &  0.053    &   3    &   3.14     \\
          &  F814W &  0.056    &   3    &   3.05     \\
WFPC2/WFC &  F300W &  0.076    &   2    &   5.05     \\
          &  F450W &  0.073    &   2    &   4.89     \\
          &  F555W &  0.064    &   2    &   4.35     \\
          &  F606W &  0.059    &   2    &   4.11     \\
          &  F814W &  0.051    &   2    &   3.71     \\
WFPC2/PC  &  F300W &  0.051    &   2    &   3.76     \\
          &  F555W &  0.045    &   2    &   3.10     \\
          &  F606W &  0.045    &   2    &   2.96     \\
          &  F814W &  0.059    &   2    &   3.18     \\
\tableline
\end{tabular}
\end{table}

\begin{sidewaystable}
\begin{center}
\small
\caption{Basic parameters of 79 sample clusters in M31.} \vspace{2mm} \label{t5.tab}
\tabcolsep=2.5pt
\begin{tabular}{cccccccccccc}
\tableline\tableline
Name & Detector  & Band  & $N_{\rm pts}$$^a$ & Model & $\chi_{\rm min}^2$$^b$ & $I_{\rm bkg}$$^c$ & $W_0$$^d$ & $c/n$$^e$ &  $\mu_0$$^f$ & $\log r_0$$^g$ & $\log r_0$$^h$ \\
     &       &    &   &      &   & ($L_{\odot}$~pc$^{-2}$) &    &    &     ${\rm (mag~arcsec^{-2})}$ & (arcsec) & (pc)\\
(1) &  (2)  &  (3) &  (4) & (5) &  (6)  & (7) &  (8)  &  (9) & (10) &  (11) &(12)  \\
\hline
 B006     & WFPC2/PC   & 555       & 49       & K66      & 35.44    & $5.10     \pm 1.54    $ & $7.90    ^{+0.14    }_{-0.21   }         $ & $1.80    ^{+0.04    }_{-0.07   }         $ & $14.43   ^{+0.04    }_{-0.07   }$ & $-0.900  ^{+0.012   }_{-0.014  }$ & $-0.320  ^{+0.012   }_{-0.014  }$\\
          &            &           & 49       & W        & 5.53     & $-8.30    \pm 0.92    $ & $7.60    ^{+0.04    }_{-0.06   }         $ & $2.86    ^{+0.03    }_{-0.05   }         $ & $14.44   ^{+0.01    }_{-0.03   }$ & $-0.850  ^{+0.009   }_{-0.007  }$ & $-0.270  ^{+0.009   }_{-0.007  }$\\
          &            &           & 49       & S        & 130.08   & $3.50     \pm 2.56    $ & $\ldots                                  $ & $2.60    ^{+0.06    }_{-0.04   }         $ & $14.28   ^{+0.12    }_{-0.08   }$ & $-2.400  ^{+0.032   }_{-0.050  }$ & $-1.820  ^{+0.032   }_{-0.050  }$\\
 B006     & WFPC2/PC   & 814       & 49       & K66      & 14.97    & $6.00     \pm 3.36    $ & $8.20    ^{+0.29    }_{-0.22   }         $ & $1.89    ^{+0.09    }_{-0.07   }         $ & $13.35   ^{+0.08    }_{-0.08   }$ & $-0.950  ^{+0.013   }_{-0.026  }$ & $-0.370  ^{+0.013   }_{-0.026  }$\\
          &            &           & 49       & W        & 1.88     & $-9.90    \pm 2.25    $ & $7.60    ^{+0.10    }_{-0.07   }         $ & $2.86    ^{+0.09    }_{-0.06   }         $ & $13.39   ^{+0.04    }_{-0.02   }$ & $-0.850  ^{+0.006   }_{-0.007  }$ & $-0.270  ^{+0.006   }_{-0.007  }$\\
          &            &           & 49       & S        & 32.24    & $4.40     \pm 3.68    $ & $\ldots                                  $ & $2.65    ^{+0.06    }_{-0.06   }         $ & $16.44   ^{+0.50    }_{-0.50   }$ & $-2.450  ^{+0.038   }_{-0.039  }$ & $-1.870  ^{+0.038   }_{-0.039  }$\\
\tableline
\end{tabular}
\end{center}
\footnotetext[1]{The number of points in the intensity profile that were used for constraining the model fits.}
\footnotetext[2]{The minimum $\chi^2$ obtained in the fits.}
\footnotetext[3]{The best-fitted background intensity.}
\footnotetext[4]{The dimensionless central potential of the best-fitting model, defined as $W_0 \equiv -\phi(0)/\sigma_0^2$.}
\footnotetext[5]{The concentration $c \equiv \log(r_t/r_0)$.}
\footnotetext[6]{The best-fit central surface brightness in the native bandpass of the data.}
\footnotetext[7]{The best model-fit scale radius $r_0$ in arcseconds.}
\footnotetext[8]{The best model-fit scale radius $r_0$ in parsecs.}
\end{sidewaystable}

\begin{sidewaystable}
\begin{center}
\small
\caption{Derived structural and photometric parameters of 79 sample clusters in M31.} \vspace{2mm} \label{t6.tab}
\tabcolsep=2.5pt
\begin{tabular}{cccccccccccccc}
\tableline\tableline
Name & Detector  & Band & Model & $\log r_{\rm tid}$$^a$ & $\log R_c$$^b$ & $\log R_h$$^c$ & $\log R_h/R_c$$^d$ & $\log I_{\rm 0}$$^e$ & $\log j_{\rm 0}$$^f$ & $\log L_V$$^g$ & $V_{\rm tot}$$^h$  & $\log I_h$$^i$ & $<\mu_V>_h$$^j$\\
     &    & &  & (pc) & (pc)  & (pc)  &   & ($L_{\odot,V}$~pc$^{-2}$) &  ($L_{\odot,V}$~pc$^{-3}$)  & ($L_{\odot,V}$)  &  (mag)   &  ($L_{\odot,V}$~pc$^{-2}$) &   ${\rm (mag~arcsec^{-2})}$  \\
(1) &  (2)  &  (3) &  (4) & (5) &  (6)  & (7) &  (8)  &  (9) & (10) &  (11) &(12) & (13) & (14) \\
\hline
 B006     & WFPC2/PC   & 555        & K66      & $1.48    ^{+0.05    }_{-0.06   }         $ & $-0.333  ^{+0.010   }_{-0.013  }$ & $0.386   ^{+0.058   }_{-0.077  }$ & $0.719   ^{+0.071   }_{-0.087  }$ & $4.79    ^{+0.03    }_{-0.03   }$ & $4.81    ^{+0.05    }_{-0.04   }$ & $5.63    ^{+0.02    }_{-0.03   }$ & $15.18   ^{+0.07    }_{-0.06   }$ & $4.06    ^{+0.12    }_{-0.09   }$ & $16.20   ^{+0.23    }_{-0.31   }$\\
          &            &            & W        & $2.59    ^{+0.04    }_{-0.04   }         $ & $-0.295  ^{+0.008   }_{-0.007  }$ & $0.511   ^{+0.046   }_{-0.086  }$ & $0.807   ^{+0.053   }_{-0.093  }$ & $4.78    ^{+0.02    }_{-0.02   }$ & $4.77    ^{+0.03    }_{-0.03   }$ & $5.71    ^{+0.02    }_{-0.03   }$ & $14.98   ^{+0.08    }_{-0.06   }$ & $3.89    ^{+0.14    }_{-0.07   }$ & $16.63   ^{+0.17    }_{-0.35   }$\\
          &            &            & S        & $\infty                                  $ & $-1.820  ^{+0.032   }_{-0.050  }$ & $0.399   ^{+0.126   }_{-0.121  }$ & $2.218   ^{+0.176   }_{-0.153  }$ & $4.85    ^{+0.04    }_{-0.05   }$ & $5.49    ^{+0.08    }_{-0.08   }$ & $5.63    ^{+0.07    }_{-0.07   }$ & $15.18   ^{+0.16    }_{-0.18   }$ & $4.04    ^{+0.18    }_{-0.18   }$ & $16.27   ^{+0.45    }_{-0.44   }$\\
 B006     & WFPC2/PC   & 814        & K66      & $1.52    ^{+0.09    }_{-0.07   }         $ & $-0.381  ^{+0.011   }_{-0.025  }$ & $0.428   ^{+0.094   }_{-0.085  }$ & $0.810   ^{+0.119   }_{-0.096  }$ & $4.79    ^{+0.03    }_{-0.03   }$ & $4.86    ^{+0.06    }_{-0.04   }$ & $5.63    ^{+0.02    }_{-0.03   }$ & $15.18   ^{+0.07    }_{-0.06   }$ & $3.98    ^{+0.14    }_{-0.16   }$ & $16.41   ^{+0.41    }_{-0.35   }$\\
          &            &            & W        & $2.59    ^{+0.09    }_{-0.06   }         $ & $-0.295  ^{+0.005   }_{-0.005  }$ & $0.513   ^{+0.093   }_{-0.044  }$ & $0.809   ^{+0.098   }_{-0.049  }$ & $4.78    ^{+0.02    }_{-0.02   }$ & $4.77    ^{+0.03    }_{-0.03   }$ & $5.71    ^{+0.02    }_{-0.03   }$ & $14.98   ^{+0.08    }_{-0.06   }$ & $3.89    ^{+0.06    }_{-0.16   }$ & $16.64   ^{+0.41    }_{-0.14   }$\\
          &            &            & S        & $\infty                                  $ & $-1.870  ^{+0.038   }_{-0.039  }$ & $0.416   ^{+0.127   }_{-0.120  }$ & $2.286   ^{+0.166   }_{-0.159  }$ & $4.85    ^{+0.04    }_{-0.05   }$ & $5.53    ^{+0.07    }_{-0.08   }$ & $5.63    ^{+0.07    }_{-0.07   }$ & $15.18   ^{+0.16    }_{-0.18   }$ & $4.00    ^{+0.18    }_{-0.18   }$ & $16.35   ^{+0.46    }_{-0.44   }$\\
\tableline
\end{tabular}
\end{center}
\footnotetext[1]{The model tidal radius $r_t$ in parsecs.}
\footnotetext[2]{The projected core radius of the model fitting a cluster, defined as $I(R_c) = I_0/2$.}
\footnotetext[3]{The projected half-light, or effective, radius of a model, containing half the total luminosity in projection.}
\footnotetext[4]{A measure of cluster concentration and relatively more model-independent than $W_0$ or $c$.}
\footnotetext[5]{The best-fit central ($R = 0$) luminosity surface density in the $V$ band, defined as $\log I_0 = 0.4(26.358 - \mu_{V,0}$), where 26.358 is the ``Coefficient'' corresponding to a solar absolute magnitude $M_{V,\odot} = +4.786$ (C. Willmer, private communication).}
\footnotetext[6]{The central ($r = 0$) luminosity volume density in the $V$ band.}
\footnotetext[7]{The $V$-band total integrated model luminosity.}
\footnotetext[8]{The total $V$-band magnitude of a model cluster, defined as $V_{\rm tot} = 4.786 - 2.5 \log (L_V/L_{\odot}) + 5\log (D/10$ pc).}
\footnotetext[9]{The luminosity surface density averaged over the half-light/effective radius in the $V$ band, defined as $\log I_h \equiv \log (L_V/2{\pi}R_h^2$).}
\footnotetext[10]{The surface brightness in magnitude over the half-light/effective radius in the $V$ band, defined as $<\mu_V>_h = 26.358 - 2.5\log I_h$.}
\end{sidewaystable}

\begin{sidewaystable}
\begin{center}
\small
\caption{Derived dynamical parameters of 79 sample clusters in M31.} \vspace{2mm} \label{t7.tab}
\tabcolsep=2.5pt
\begin{tabular}{cccccccccccccc}
\tableline\tableline
Name & Detector & Band &  $\Upsilon_V^{\rm pop}$$^a$  & Model & $\log M_{\rm tot}$$^b$ & $\log E_b$$^c$ & $\log \Sigma_{\rm 0}$$^d$ & $\log \rho_{\rm 0}$$^e$ & $\log \Sigma_h$$^f$ & $\log \sigma_{p,0}$$^g$ & $\log \nu_{\rm esc,0}$$^h$ & $\log t_{r,h}$$^i$ & $\log f_{\rm 0}$$^j$\\
     &  &   & ($M_{\odot}~L_{\odot,V}^{-1}$)  & & ($M_{\odot}$) & (erg)  & ($M_{\odot}$~pc$^{-2}$) & ($M_{\odot}$~pc$^{-3}$)   &  ($M_{\odot}$~pc$^{-2}$) &  (km s$^{-1}$)  & (km s$^{-1}$)  & (yr) & ($M_{\odot}$~(pc~km~s$^{-1})^{-3})$ \\
(1) &  (2)  &  (3) &  (4) & (5) &  (6)  & (7) &  (8)  &  (9) & (10) &  (11) &(12) & (13)  & (14) \\
\hline
 B006     & WFPC2/PC   & 555        & $2.588   ^{+0.810   }_{-0.565  }         $ & K66      & $6.05    ^{+0.12    }_{-0.11   }$ & $51.76   ^{+0.48    }_{-0.43   }$ & $5.20    ^{+0.12    }_{-0.11   }$ & $5.23    ^{+0.13    }_{-0.11   }$ & $4.48    ^{+0.15    }_{-0.16   }$ & $1.172   ^{+0.060   }_{-0.054  }$ & $1.783   ^{+0.061   }_{-0.055  }$ & $9.08    ^{+0.15    }_{-0.17   }$ & $0.493   ^{+0.062   }_{-0.056  }$\\
          &            &            & $                                        $ & W        & $6.12    ^{+0.12    }_{-0.11   }$ & $51.81   ^{+0.47    }_{-0.43   }$ & $5.20    ^{+0.12    }_{-0.11   }$ & $5.19    ^{+0.12    }_{-0.11   }$ & $4.30    ^{+0.14    }_{-0.18   }$ & $1.190   ^{+0.060   }_{-0.054  }$ & $1.803   ^{+0.060   }_{-0.054  }$ & $9.30    ^{+0.14    }_{-0.18   }$ & $0.387   ^{+0.059   }_{-0.054  }$\\
          &            &            & $                                        $ & S        & $6.05    ^{+0.14    }_{-0.13   }$ & $50.39   ^{+0.48    }_{-0.44   }$ & $5.26    ^{+0.12    }_{-0.12   }$ & $5.90    ^{+0.14    }_{-0.13   }$ & $4.45    ^{+0.22    }_{-0.21   }$ & $0.714   ^{+0.059   }_{-0.056  }$ & $1.480   ^{+0.059   }_{-0.054  }$ & $9.10    ^{+0.25    }_{-0.24   }$ & $3.054   ^{+0.111   }_{-0.070  }$\\
 B006     & WFPC2/PC   & 814        & $2.588   ^{+0.810   }_{-0.565  }         $ & K66      & $6.05    ^{+0.12    }_{-0.11   }$ & $51.66   ^{+0.47    }_{-0.43   }$ & $5.20    ^{+0.12    }_{-0.11   }$ & $5.28    ^{+0.13    }_{-0.11   }$ & $4.39    ^{+0.20    }_{-0.18   }$ & $1.148   ^{+0.059   }_{-0.054  }$ & $1.765   ^{+0.060   }_{-0.055  }$ & $9.14    ^{+0.19    }_{-0.18   }$ & $0.617   ^{+0.076   }_{-0.056  }$\\
          &            &            & $                                        $ & W        & $6.12    ^{+0.12    }_{-0.11   }$ & $51.81   ^{+0.48    }_{-0.43   }$ & $5.20    ^{+0.12    }_{-0.11   }$ & $5.19    ^{+0.12    }_{-0.11   }$ & $4.30    ^{+0.20    }_{-0.12   }$ & $1.190   ^{+0.060   }_{-0.054  }$ & $1.803   ^{+0.060   }_{-0.054  }$ & $9.30    ^{+0.19    }_{-0.13   }$ & $0.387   ^{+0.059   }_{-0.054  }$\\
          &            &            & $                                        $ & S        & $6.05    ^{+0.14    }_{-0.13   }$ & $50.36   ^{+0.49    }_{-0.45   }$ & $5.26    ^{+0.12    }_{-0.12   }$ & $5.94    ^{+0.14    }_{-0.13   }$ & $4.42    ^{+0.22    }_{-0.21   }$ & $0.698   ^{+0.060   }_{-0.057  }$ & $1.470   ^{+0.059   }_{-0.054  }$ & $9.13    ^{+0.25    }_{-0.23   }$ & $3.168   ^{+0.089   }_{-0.078  }$\\
\tableline
\end{tabular}
\end{center}
\footnotetext[1]{The $V$-band mass-to-light ratio.}
\footnotetext[2]{The integrated cluster mass, estimated as $\log M_{\rm tot} = \log \Upsilon_V^{\rm pop} + \log L_V$.}
\footnotetext[3]{The integrated binding energy in ergs, defined as $E_b \equiv -(1/2) \int_0^{r_t} 4{\rm \pi}r^2\rho\phi{\rm d}r$.}
\footnotetext[4]{The central surface mass density, estimated as $\log \Sigma_0=\log \Upsilon_V^{\rm pop} + \log I_0$.}
\footnotetext[5]{The central volume density, estimated as $\log \rho_0 = \log \Upsilon_V^{\rm pop} + \log j_0$.}
\footnotetext[6]{The surface mass density averaged over the half-light/effective radius $R_h$, estimated as $\log \Sigma_h = \log \Upsilon_V^{\rm pop} + \log I_h$.}
\footnotetext[7]{The predicted line-of-sight velocity dispersion at the cluster center.}
\footnotetext[8]{The predicted central ``escape'' velocity with which a star can move out from the center of a cluster, defined as $\nu_{\rm esc,0}^2/\sigma_0^2 = 2[W_0 + GM_{\rm tot}/r_t\sigma_0^2]$.}
\footnotetext[9]{The two-body relaxation time at the model-projected half-mass radius, estimated as $t_{r,h} = {\frac{2.06\times10^6 yr}{\ln(0.4M_{\rm tot}/m_{\star})}}{\frac{M_{\rm tot}^{1/2}R_h^{3/2}}{m_{\star}}}$. $m_{\star}$ is the average stellar mass in a cluster, assumed to be 0.5$M_{\odot}$.}
\footnotetext[10]{The model's central phase-space density, defined as $\log f_0 \equiv \log [\rho_0/(2\pi\sigma_c^2)^{3/2}$].}
\end{sidewaystable}


\begin{thebibliography}{}

\bibitem[Baes \& Gentile(2011)]{bg11}Baes, M. \& Gentile, G. 2011, A\&A, 525, A136

\bibitem[Barmby et al.(2002)]{bhh02}Barmby, P., Holland, S., \& Huchra, J. P. 2002, AJ, 123, 1937

\bibitem[Barmby et al.(2000)]{bh00}Barmby, P., Huchra, J., Brodie, J., et al. 2000, AJ, 119, 727

\bibitem[Barmby et al.(2007)]{barmby07}Barmby, P., McLaughlin, D. E., Harris, W. E., Harris, G. L. H., \& Forbes, D. A. 2007, AJ, 133, 2764

\bibitem[Barmby et al.(2009)]{barmby09}Barmby, P., Perina, S., Bellazzini, M., et al. 2009, AJ, 138, 1667

\bibitem[Battistini et al.(1982)]{battistini82}Battistini, P., Bonoli, F., Pecci, F.~F., Buonanno, R., \& Corsi, C.~E.\ 1982, \aap, 113, 39

\bibitem[Bekki(2010)]{bekki10}Bekki, K. 2010, MNRAS, 401, L58

\bibitem[Bellazzini et al.(2003)]{bfi}Bellazzini, M., Ferraro, F. R., \& Ibata, R. A. 2003, \aj, 125, 188


\bibitem[Bendinelli et al.(1993)]{bendinelli93}Bendinelli, O., Cacciari, C., Djorgovski, S., et al.\ 1993, \apjl, 409, L17

\bibitem[Bendinelli et al.(1990)]{bendinelli90}Bendinelli, O., Zavatti, F., Parmeggiani, G., \& Djorgovski, S.\ 1990, \aj, 99, 774

\bibitem[Bernard et al.(2012)]{bernard12}Bernard, E.~J., Ferguson, A.~M.~N., Barker, M.~K., et al.\ 2012, \mnras, 420, 2625




\bibitem[Bruzual \& Charlot(2003)]{bc03}Bruzual, G., \& Charlot, S. 2003, MNRAS, 344, 1000

\bibitem[Caldwell et al.(2009)]{cald09}Caldwell, N., Harding, P., Morrison, H., et al. 2009, AJ, 137, 94

\bibitem[Caldwell et al.(2011)]{cald11}Caldwell, N., Schiavon, R., Morrison, H., Rose, J. A., \& Harding, P. 2011, AJ, 141, 61

\bibitem[Cardelli et al.(1989)]{car89}Cardelli, J. A., Clayton, G. C., \& Mathis, J. S. 1989, ApJ, 345, 245

\bibitem[Chabrier(2003)]{chab03}Chabrier, G. 2003, PASP, 115, 763

\bibitem[Chatterjee et al.(2012)]{chatt12}Chatterjee, S., Umbreit, S., Fregeau, J.~M., \& Rasio, F.~A.\ 2012, arXiv:1207.3063

\bibitem[Cohen \& Freeman(1991)]{cf91}Cohen, J.~G., \& Freeman, K.~C.\ 1991, \aj, 101, 483

\bibitem[Crampton et al.(1985)]{crampton85}Crampton, D., Cowley, A.~P., Schade, D., \& Chayer, P.\ 1985, \apj, 288, 494



\bibitem[Davoust \& Prugniel(1990)]{dp90}Davoust, E., \& Prugniel, P. 1990, A\&A, 230, 67

\bibitem[Da Costa \& Freeman(1976)]{df76}Da Costa, G.~S., \& Freeman, K.~C.\ 1976, \apj, 206, 128

\bibitem[Djorgovski(1995)]{djor95}Djorgovski, S.\ 1995, \apjl, 438, L29

\bibitem[Djorgovski \& Meylan(1994)]{dm94}Djorgovski, S., \& Meylan, G.\ 1994, \aj, 108, 1292

\bibitem[Dull et al.(1997)]{dull97}Dull, J.~D., Cohn, H.~N., Lugger, P.~M., et al.\ 1997, \apj, 481, 267

\bibitem[Elson, Fall \& Freeman(1987)]{elson87}Elson, R. A. W., Fall, S. M., \& Freeman, K. C. 1987, ApJ, 323, 54


\bibitem[Federici et al.(2007)]{federici07}Federici, L., Bellazzini, M., Galleti, S., et al. 2007, A\&A, 473, 429

\bibitem[Fusi Pecci et al.(1994)]{fusi94}Fusi Pecci, F., Battistini, P., Bendinelli, O., et al.\ 1994, \aap, 284, 349

\bibitem[Galleti et al.(2009)]{gall09}Galleti, S., Bellazzini, M., Buzzoni, A., Federici, L., \& Fusi Pecci, F. 2009, A\&A, 508, 1285

\bibitem[Galleti et al.(2006)]{gall06}Galleti, S., Federici, L., Bellazzini, M., Buzzoni, A., \& Fusi Pecci, F. 2006, A\&A, 456, 985

\bibitem[Galleti et al.(2004)]{gall04}Galleti, S., Federici, L., Bellazzini, M., Fusi Pecci, F., \& Macrina, S. 2004, A\&A, 426, 917

\bibitem[Georgiev et al.(2009)]{Georgiev09}Georgiev, I. Y., Hilker, M., Puzia, T. H., Goudfrooij, P., \& Baumgardt, H. 2009, MNRAS, 396, 1075

\bibitem[Geyer et al.(2009)]{Geyer83}Geyer, E. H., Nelles, B., \& Hopp, U. 1983, A\&A, 125, 359

\bibitem[Gnedin et al.(1999)]{glo99}Gnedin, O. Y., Lee,
H. M., \& Ostriker, J. P. 1999, \apj, 522, 935

\bibitem[Goodwin(1997)]{goodwin97}Goodwin, S.~P.\ 1997, \mnras, 286, L39


\bibitem[Grillmair et al.(1996)]{grill96}Grillmair, C.~J., Ajhar, E.~A., Faber, S.~M., et al.\ 1996, \aj, 111, 2293

\bibitem[Gunn \& Griffin(1979)]{gg79}Gunn, J.~E., \& Griffin, R.~F.\ 1979, \aj, 84, 752


\bibitem[Harris(1996)]{harris96}Harris, W. E. 1996, AJ, 112, 1487

\bibitem[Harris et al.(2002)]{harris02}Harris, W. E., Harris, G. L. H., Holland, S. T., \& McLaughlin, D. E. 2002, AJ, 124, 1435

\bibitem[Hill \& Zaritsky(2006)]{hz06}Hill, A., \& Zaritsky, D.\ 2006, \aj, 131, 414


\bibitem[Holtzman et al.(1995)]{holtzman95}Holtzman, J.~A., Burrows, C.~J., Casertano, S., et al.\ 1995, \pasp, 107, 1065

\bibitem[Huchra et al.(1991)]{huchra91}Huchra, J. P., Brodie, J. P., \& Kent, S. M. 1991, ApJ, 370, 495

\bibitem[Huxor et al.(2011)]{huxor11}Huxor, A. P., Ferguson, A. M. N., Tanvir, N. R., et al. 2011, MNRAS, 414, 770


\bibitem[Kang et al.(2012)]{kang12}Kang, Y., Rey, S.-C., Bianchi, L., et al. 2012, ApJS, 199, 37

\bibitem[King(1962)]{king62}King, I. R.\ 1962, \aj, 67, 471

\bibitem[King(1966)]{king66}King, I. R. 1966, AJ, 71, 64

\bibitem[Krist et al.(2011)]{krist11}Krist, J. E., Hook, R. N., \& Stoehr, F. 2011, in Optical Modeling and Performance Predictions V, Proc. of SPIE Vol. 8127, ed. M. A. Kahan, 81270J

\bibitem[Larsen et al.(2001)]{larsen01}Larsen, S. S., AJ, 122, 1782


\bibitem[Larsen et al.(2002)]{larsen02}Larsen, S. S., Brodie, J. P., Sarajedini, A., \& Huchra, J. P. 2002, AJ, 124, 2615


\bibitem[Ma et al.(2007)]{ma07}Ma, J., de Grijs, R., Chen, D., et al. 2007, MNRAS, 376, 1621

\bibitem[Ma et al.(2006)]{ma06}Ma, J., van den Bergh, S., Wu, H., et al. 2006, ApJ, 636, L93

\bibitem[Ma et al.(2012)]{ma12}Ma, J., Wang, S., Wu, Z., et al. 2012, AJ, 143, 29

\bibitem[Mackey \& Gilmore(2003)]{mg03}Mackey, A. D., \& Gilmore, G. F. 2003, MNRAS, 338, 85



\bibitem[Mackey et al.(2010)]{mackey10}Mackey, A.~D., Huxor, A.~P., Ferguson, A.~M.~N., et al.\ 2010, \apjl, 717, L11

\bibitem[Mackey et al.(2006)]{mackey06}Mackey, A.~D., Huxor, A., Ferguson, A.~M.~N., et al.\ 2006, \apjl, 653, L105

\bibitem[Mackey et al.(2007)]{mackey07}Mackey, A.~D., Huxor, A., Ferguson, A.~M.~N., et al.\ 2007, \apjl, 655, L85


\bibitem[McConnachie et al.(2009)]{McConnachie09}McConnachie, A.~W., Irwin, M.~J., Ibata, R.~A., et al.\ 2009, \nat, 461, 66

\bibitem[McLaughlin(2000)]{mclau00}McLaughlin, D.~E.\ 2000, \apj, 539, 618

\bibitem[Mclaughlin et al.(2008)]{mclau08}McLaughlin, D. E., Barmby, P., Harris, W. E., Forbes, D. A., \& Harris, G. L. H. 2008, MNRAS, 384, 563

\bibitem[Mclaughlin \& van der Marel(2005)]{mm05}McLaughlin, D. E., \& van der Marel, R. P. 2005, ApJS, 161, 304




\bibitem[Meylan(1988)]{meylan88}Meylan, G.\ 1988, \aap, 191, 215

\bibitem[Meylan(1989)]{meylan89}Meylan, G.\ 1989, \aap, 214, 106

\bibitem[Meylan \& Heggie(1997)]{mh97}Meylan, G., \& Heggie, D.~C. \ 1997, \aapr, 8, 1

\bibitem[Noyola \& Gebhardt(2006)]{ng06}Noyola, E. \& Gebhardt, K. 2006, AJ, 132, 447

\bibitem[Perina et al.(2010)]{perina10}Perina, S., Cohen, J.~G., Barmby, P., et al.\ 2010, \aap, 511, A23


\bibitem[Portegies et al.(2010)]{portegies10}Portegies Zwart, S., McMillan, S., \& Gieles, M. 2010, ARA\&A, 48, 431

\bibitem[Pritchet \& van den Bergh(1984)]{pv84}Pritchet, C., \& van den Bergh, S.\ 1984, \pasp, 96, 804

\bibitem[Putman et al.(2009)]{putman09}Putman, M.~E., Peek, J.~E.~G., Muratov, A., et al.\ 2009, \apj, 703, 1486




\bibitem[Rhodes et al.(2006)]{rhodes06}Rhodes, J. D., Massey, R., Albert, J., et al. 2006, in The 2005 HST Calibration Workshop: Hubble After the Transition to Two-Gyro Mode, ed. A. M. Koekemoer, P. Goudfrooij, \& L. L. Dressel, 21

\bibitem[Richardson et al.(2009)]{rich09}Richardson, J.~C., Ferguson, A.~M.~N., Mackey, A.~D., et al.\ 2009, \mnras, 396, 1842

\bibitem[Saito(1979)]{saito79}Saito, M.\ 1979, \pasj, 31, 181

\bibitem[San Roman et al.(2010)]{roman10}San Roman, I., Sarajedini, A., \& Aparicio, A.\ 2010, \apj, 720, 1674

\bibitem[San Roman et al.(2012)]{roman12}San Roman, I., Sarajedini, A., Holtzman, J.~A., \& Garnett, D.~R.\ 2012, \mnras, 426, 2427

\bibitem[S\'{e}rsic(1968)]{sersic68}S\'{e}rsic, J.-L. 1968, Atlas de Galaxias Australes (Cordoba: Obs. Astronomico)

\bibitem[Sippel et al.(2012)]{sippel12}Sippel, A. C., Hurley, J. R., Madrid, J. P.,  \& Harris, W. E. 2012, MNRAS, 427, 167

\bibitem[Sirianni et al.(2005)]{siri05}Sirianni, M., Jee, M. J., Ben\`{\i}tez, N. et al. 2005, PASP, 117, 1049

\bibitem[Spassova et al.(1988)]{spassova88}Spassova, N.~M., Staneva, A.~V., \& Golev, V.~K.\ 1988, The Harlow-Shapley Symposium on Globular Cluster Systems in Galaxies, 126, 569

\bibitem[Stanek \& Garnavich(1998)]{sg98}Stanek, K.~Z., \& Garnavich, P.~M.\ 1998, \apjl, 503, L131

\bibitem[Strader et al.(2011)]{str11}Strader, J., Caldwell, N., \& Seth, A. C. 2011, AJ, 142, 8

\bibitem[Strader et al.(2012)]{str12}Strader, J., Fabbiano, G., Luo, B., et al. 2012, ApJ, 760, 87

\bibitem[Strader et al.(2009)]{str09}Strader, J., Smith, G.~H., Larsen, S., Brodie, J.~P., \& Huchra, J.~P.\ 2009, \aj, 138, 547

\bibitem[Tanvir et al.(2012)]{tanvir12}Tanvir, N. R., Mackey, A. D., Ferguson, A. M. N., et al. 2012, MNRAS, 422, 162


\bibitem[Trager et al.(1995)]{tkd95}Trager, S. C., King, I. R., \& Djorgovski, S. 1995, AJ, 109, 218

\bibitem[van den Bergh(1969)]{vanden69}van den Bergh, S. 1969, ApJS, 19, 145

\bibitem[van den Bergh(1991)]{vanden91}van den Bergh, S. 1991, PASP, 103, 1053

\bibitem[van den Bergh(1996)]{vanden96}van den Bergh, S. 1996, Observatory, 116, 103

\bibitem[van den Bergh \& Morbey(1984)]{vanden84}van den Bergh, S., \& Morbey, C. L. 1984, ApJ, 283, 598

\bibitem[Wang et al.(2010)]{wang10}Wang, S., Fan, Z., Ma, J., de Grijs, R., \& Zhou, X.\ 2010, \aj, 139, 1438

\bibitem[Wang \& Ma(2012)]{wang12a}Wang, S., \& Ma, J. 2012, AJ, 143, 132

\bibitem[Wang et al.(2012)]{wang12b}Wang, S. Ma, J. Fan, Z. Wu, Z., Zhang, T., Zou, H., \& Zhou, X.\ 2012, \aj, 144, 191

\bibitem[Werchan \& Zaritsky(2011)]{wz11}Werchan, F., \& Zaritsky, D.\ 2011, \aj, 142, 48

\bibitem[Wilson(1975)]{wilson75}Wilson, C. P. 1975, AJ, 80, 175
\end{thebibliography}
\end{document}